\documentclass[twocolumn,prc,aps,showpacs,preprintnumbers,amsmath,amssymb]{revtex4}

\usepackage{graphicx}

 \newcommand\tr{{\rm Tr\,}}
 \newcommand\la{\langle}
 \newcommand\ra{\rangle}
 \newcommand\beq{\begin{equation}}
 
 \newcommand\eeq{\end{equation}}
 \newcommand\beqn{\begin{eqnarray}}
 \newcommand\eeqn{\end{eqnarray}}

\def\mub{\,\mu\mbox{b}}
\def\fm{\,\mbox{fm}}
\def\GeV{\,\mbox{GeV}}

\def\Pom{{\bf I\!P}}

\def\lsim{\mathrel{\rlap{\lower4pt\hbox{\hskip1pt$\sim$}}
    \raise1pt\hbox{$<$}}}         
\def\gsim{\mathrel{\rlap{\lower4pt\hbox{\hskip1pt$\sim$}}
    \raise1pt\hbox{$>$}}}         

\begin{document}

\title{Diffractive Excitation of Heavy Flavors:
Leading Twist Mechanisms}

\author{B.Z.~Kopeliovich$^{a,b}$}
\author{I.K.~Potashnikova$^{a,b}$}
\author{Ivan~Schmidt$^a$}
\author{A.V.~Tarasov$^{b}$}

\affiliation{$^a$Departamento de F\'\i sica
y Centro de Estudios
Subat\'omicos,\\ Universidad T\'ecnica
Federico Santa Mar\'\i a, Casilla 110-V, Valpara\'\i so, Chile\\
$^b$Joint Institute for Nuclear Research, Dubna, Russia}

\date{\today}

\begin{abstract}

Diffractive production of heavy flavors is calculated within the
light-cone dipole approach. Novel leading twist mechanisms are
proposed, which involve both short and long transverse distances
inside the incoming hadron. Nevertheless, the diffractive cross
section turns out to be sensitive to the primordial transverse
momenta of projectile gluons, rather than to the hadronic size.
Our calculations agree with the available data for diffractive
production of charm and beauty, and with the observed weak
variation of the diffraction-to-inclusive cross section ratios as
function of the hard scale.

\end{abstract}


\maketitle

\section{Introduction}

Diffraction is usually viewed as shadow of inelastic processes. This idea
originated from optics helped considerably in the interpretation of data on
elastic hadronic scattering. The understanding of the mechanisms of
inelastic diffraction came with the pioneering works of Glauber
\cite{glauber}, Feinberg and Pomeranchuk \cite{f-p}, Good and Walker
\cite{g-w}. If the incoming plane wave contains components interacting
differently with the target, the outgoing wave will have a different
composition, i.e. besides elastic scattering a new state will be created
(e.g. see in \cite{brazil}). An optical analogy for inelastic diffraction
would be a change of the light polarization after passing through a
polarizer which absorbs differently two linear polarizations.

Thus, inelastic diffraction of hadrons should also be treated as a shadow,
it emerges due to difference of shadows produced by different inelastic
processes (in this respect a study of the structure function \cite{i-s} of a
shadow might look questionable). Some of the shadows from different hadronic
components are large, and this gives rise to the main bulk of soft
diffractive events. However, the difference between shadows of two hadronic
components which differ from each other only by either the presence or
absence of a hard fluctuation (heavy flavors, high $p_T$ partons, heavy
dileptons, etc.) should be vanishingly small. This small difference gives
rise to hard diffraction. It can be as small as $1/Q^2$, where $Q$ is the
hard scale, and then the amplitude squared leads to higher twist terms in
the diffractive cross section. However, in a non-abelian theory like QCD the
difference between shadow amplitudes may be larger, $\sim 1/Q$, resulting in
a leading twist behavior. 

A proper example is the leading twist diffractive Drell-Yan reaction
\cite{kpst-dy}, where simultaneously large and small size projectile
fluctuations are at work. Due to general properties of diffraction a single
quark cannot radiate diffractively a photon or a dilepton (or any colorless and
point-like particle) in forward direction (when the recoil target proton has
transverse momentum $p_T=0$) \cite{kst2}. Let us consider two Fock components of
a physical quark, a bare quark $|q\ra$, and a quark accompanied by a photon,
$|q\gamma^*\ra$.  In both Fock states only the quark is able to interact.  
Therefore, the forward diffractive amplitude given by the difference of these
two elastic amplitude integrated over impact parameter, vanishes.

Nevertheless, a $\bar qq$ dipole (or any hadron) can radiate diffractively in 
forward direction. Indeed, in this case the two Fock states $|\bar qq\ra$ and 
$|\bar qq\gamma^*\ra$ have different cross sections and the difference is,
 \beqn
A_{diff}(\bar qq+p\to\bar qq\gamma^*+p)|_{p_T=0}
 &\propto&
\sigma_{\bar qq}(R)-\sigma_{\bar qq}(\vec R+\vec r)
\nonumber \\
&\propto& \vec r\cdot\vec R +O(r^2).
\label{01}
 \eeqn
 Here $R$ is the $\bar qq$ transverse separation; $r\sim 1/Q$ is the small shift
of impact parameter of the quark radiating the heavy photon; $\sigma_{\bar
qq}(R)$ is the dipole-proton cross section which we assume for simplicity to be
quadratic in $R$. It is demonstrated in \cite{kpst-dy} that integration over
azimuthal angle does not terminate the leading term $(\vec r\cdot\vec R)$ due to
the convolution with the final state wave function. Thus, the amplitude has a
leading twist scale dependence, $1/Q$.

The origin of leading twist behavior of heavy flavor diffractive production
has some similarities, but also differences. In this paper we classify different
mechanisms of diffractive excitation of heavy flavors, with the purpose of
identifying leading twist terms. It is instructive to compare them with the well
studied case of diffractive deep-inelastic scattering (DIS).

\subsection{Diffractive production of heavy quarks in DIS}

The fraction of DIS events with diffractive excitation of the virtual photon
has been found in experiments at HERA to be nearly scale independent
\cite{hera}. This might have been a surprise, since diffraction, which is a
large rapidity gap (LRG) process, is associated with a Pomeron, i.e. at
least two gluon exchange. If each of the gluons has to resolve a hard photon
fluctuation, $\gamma^*\to\bar qq$, of a size $1/Q^2$ ($Q$ is the photon
virtuality), then to do it twice costs more and should lead to a cross
section as small as $1/Q^4$. However, such an expectation contradicts data.
It turns out that in this case the main contribution for transversely
polarized photons comes from rare fluctuations of the photon. These
fluctuations correspond to aligned jet configurations \cite{bjorken} which
happen rarely, but are soft and interact strongly \cite{kp}.

Although the cross section of diffractive dissociation
$\gamma^*\to\bar qq$ seems to behave as leading twist,
$\sigma_{sd}(\gamma^*p\to\bar qqp\sim 1/Q^2)$, in fact this is a
higher twist process. Indeed, if we impose the hard scale to be
the heavy quark mass $m_Q^2\gg Q^2$, the same process behaves as
one would expect for a higher twist,
 \beq
\sigma_{sd}(\gamma^*+p\to\bar QQ+p)\propto 1/m_Q^4.
 \label{02}
 \eeq

The real leading twist behavior emerges from more complicated
photon fluctuations which contain at least one gluon besides the
$\bar QQ$, $\gamma^*\to\bar QQG$ (e.g. see in \cite{kst2,levin-diff}).
Such fluctuations are characterized by two sizes, one controlled
by the hard scale, which is small, either $r\sim 1/Q^2$ or
$r\sim1/m_Q^2$. Another size, the mean quark-gluon separation, is
large and depends only logarithmically on the
scale. This fact gives rise to the leading twist behavior of
diffraction: one of the $t$-channel gluons has to resolve the
small ($1/Q^2$ or $1/m_Q^2$) size, while another gluon may
interact with the large $\bar QQ-G$ dipole. This corresponds to
diffractive excitation, which is different from (\ref{02}),
 \beq
\sigma_{sd}(\gamma^*+p\to\bar QQg+p)\propto 1/m_Q^2. \label{04}
 \eeq
 Gluon radiation in the final state is here essential for the leading twist
behavior. Indeed, although higher Fock components, like $\bar QQg$, also
contribute to (\ref{02}), but this does not not change its higher twist scale
dependence.

Notice that diffraction is closely related to nuclear shadowing,
since both emerge from the unitarity relation as a shadow of
inelastic processes. A direct relation between diffractive
excitations of the beam and nuclear shadowing was first found in
\cite{gribov}, and is known as Gribov inelastic shadowing. In our
case, the leading and higher twist contributions to shadowing are
related to the same types of fluctuations of the photon, $\bar
QQG...$ and $\bar QQ$ respectively \cite{krt1,krt2}.

\subsection{Diffractive hadroproduction of heavy quarks}

One may treat diffraction in DIS as a way to measure the partonic
structure of the Pomeron \cite{i-s}. Having this kind of
information one may try to predict other hard diffractive
processes assuming factorization. However, attempts to apply QCD
factorization to hard diffraction failed by an order of magnitude
when one compares diffraction in DIS and in hadronic collisions
\cite{tevatron}.

There are many reasons for this breakdown of factorization. The
first one is pretty obvious, and comes from the absorptive or
unitarity corrections, which has been known since the era of Regge
phenomenology. These effects cause the suppression of any LRG
process, except elastic scattering. In the limit of unitarity
saturation (black disk) the absorptive corrections may completely
terminate the LRG process. Actually, this almost happens in
(anti)proton-proton collisions, where unitarity is nearly
saturated at small impact parameters \cite{k3p}. The suppression
factor, which is also called survival probability, changes the
diffractive cross section by an order of magnitude. Although hard
reactions hardly make any shadow, the strength of absorptive
corrections in hadronic collisions is controlled by the soft
spectator partons (see Sect.~\ref{survival}, which are absent in
the case of diffraction in DIS. This is why factorization is
severely broken.

Another source of factorization breaking is the difference between
the mechanisms of diffractive  $\bar QQ$ pair production in DIS
and in hadronic collisions. In both cases the Pomeron (i.e. two or
more gluons) can be attached directly to the produced heavy
quarks, and this part of the interaction is subject to
factorization. In a hadronic collision, however, the Pomeron can
be attached simultaneously to the projectile gluon and to the
heavy quarks. In other words, the heavy pair, which has a lifetime
substantially shorter than the projectile gluon in the incoming
hadron, may be produced during the interaction \cite{levin-coh}.
This part, called coherent diffraction \cite{cfs}, causes another
deviation from factorization.

This part of diffraction was modeled in \cite{levin-coh} by a colorless
projectile gluon diffractively dissociating into $\bar QQ$. The cross
section was found to be leading twist, $\propto 1/m_Q^2$. The confusion
caused by the graphic presentation (Figs.~3 and 5 of \cite{levin-coh})
motivated a more complete calculation in \cite{yuan-chao} of the full set
of Feynman graphs corresponding to diffractive dissociation of a colored
gluon, $g+p\to\bar QQ+p$. These authors found this process to be a higher
twist, like in photoproduction Eq.~(\ref{02}). Indeed, since in the
dissociation $g\to\bar QQ$ all transverse distances between the qluon and
quarks are of the order of $1/m_Q$, the cross section of this process
mediated by Pomeron exchange must be $\propto 1/m_Q^4$. Then this part of
diffraction violating factorization has the same $m_Q$ dependence as the
factorized one, i.e. both are higher twists.

A new source for breakdown of factorization was found in \cite{kpst-dy}
for diffractive Drell-Yan processes.  It turns out that in this case the
participation of soft spectator partons in the interaction with the
Pomeron is crucial and results in a leading twist effect. A similar
mechanism for hadroproduction of heavy quarks is under consideration in
the present paper.  It is related to the processes,
 \beqn
\sigma_{sd}(q+p\to q\bar QQ+p)&\propto& 1/m_Q^2. \label{06}\\
\sigma_{sd}(g+p\to g\bar QQ+p)&\propto& 1/m_Q^2. \label{08}
 \eeqn
 Just as in leading twist diffraction in DIS, Eq.~(\ref{04}), these
processes are associated with two characteristic transverse
separations, a small one, $\sim1/m_Q$, between the $Q$ and $\bar
Q$, and a large one, either $\sim 1/m_q$ between $q$ and $\bar QQ$
in (\ref{06}), or $\sim 1/m_g$ between $g$ and $\bar QQ$ in
(\ref{08}). Here $m_q\sim\Lambda_{QCD}$ and the several times
larger $m_g$ (see \cite{k3p,spots} are the effective cut-off
parameters which take care of the nonperturbative interactions of
quarks and gluons respectively.

Somewhat similar, but nevertheless different observations were made in Ref.
\cite{w-m}. Namely, similar to Drell-Yan diffraction \cite{kpst-dy} the large
hadronic size enters the leading twist diffractive amplitude due to the
interaction of the projectile remnants with the target. There it was concluded
that theoretical predictions cannot be certain since we are lacking reliable
information about the hadronic wave function. However, the main leading twist
contributions Eqs.~(\ref{06})-(\ref{08}) under consideration in the present
paper are independent of the structure of the incoming hadron. They correspond
to diffractive excitation of an individual parton via the so called production
mechanism (see Sect.~\ref{inclusive}).

Interaction with spectators is also known \cite{bhs} to cause considerable
effects in the azimuthal single-spin asymmetry in semi-inclusive pion
leptoproduction. In this case the outgoing quark experiences final state
interactions with remnants of the proton.

\subsection{Intrinsic heavy flavors in the proton}

Production of heavy flavors at large Feynman $x_F$ has been always
a controversial issue, even in the simple case of inclusive
processes. In the perturbative QCD approach based on QCD
factorization, inclusive heavy quark production is described as
glue-glue fusion, and the rapidity distribution of produced heavy
flavored hadrons is controlled by the gluon distribution in the
colliding hadrons. The gluon density steeply vanishes towards
$x_F=1$, approximately as $(1-x_F)^5$. Convolution with the
fragmentation function (assuming factorization) makes this
behavior at $x_F\to 1$ even steeper. On the other hand, the
end-point behavior is dictated by the general result of Regge
asymptotics,
 \beq
\left.\frac{d\sigma}{dx_F\,dt}\right|_{x_F\to1}\propto
(1-x_F)^{1-2\alpha_R(t)}\ , \label{5}
 \eeq
 where $\alpha_R(t)$ is the Regge trajectory corresponding to the
$t$-channel exchange of a heavy flavored meson or baryon, which
depends on the quantum numbers of the projectile and the produced
heavy flavored hadron. Apparently this has little to do with the
gluon distribution function. The same problem appears in the
Drell-Yan reaction at $x_F\to 1$, as is seen in data \cite{holt}.
Therefore, one should not rely on QCD factorization at large
$x_F\to1$. In fact, in this kinematic region several mechanisms
breaking factorization are known \cite{jknps}.

An excess of heavy flavored hadrons at large $x_F$ is expected to
be an evidence for the presence of intrinsic heavy flavors in the
projectile hadron \cite{stan,pumplin}. Notice, however, that any
mechanism must comply with the Regge behavior of Eq.~(\ref{5})
independently of the details of hadronic structure, i.e. must be
the same with or without the presence of intrinsic heavy flavors.
Besides, to be confident that an excess of heavy flavor is
observed, one must be able to provide a reliable theoretical
prediction for the conventional mechanisms at large $x_F$. This is
a difficult task in the situation when QCD factorization is
broken.

The observation of diffractive production of heavy quarks may provide a better
evidence for intrinsic heavy flavors. This is expected to be seen as an excess
of diffractive production compared to the conventional expectation. The latter,
therefore, must be reliably known. However, an observed signal might be
misinterpreted if it is compared with calculations assuming that factorization
holds for hard diffraction \cite{i-s,landshoff}. This is not correct as was
explained above and is confirmed by direct calculations shown below. A good
signature for the contribution of intrinsic heavy flavors would be the sharing
of longitudinal momentum in the diffractive excitation.  The intrinsic heavy
quarks should carry the main part of the momentum, which would be unusual for
the conventional mechanisms (see Sect.~\ref{results}). However, this requires
the observation of both heavy quarks, which is difficult.

Notice that diffractive production of heavy flavors also creates a
large background for searches of Higgs bosons, which can be
produced at large $x_F$ from intrinsic heavy flavors \cite{bkss}.
Moreover, one also needs to know the rate of direct production of
heavy flavors while searching for physics beyond the standard
model.

\subsection{Outline of the paper}

We present below a calculation for diffractive production of heavy
quarks in proton-proton collisions. For this purpose we rely on
the dipole approach, which is an alternative phenomenology to the
parton model formalism based on QCD factorization. The key
ingredient is the universal dipole cross section introduced in
\cite{zkl}, which is the total cross section, $\sigma_{\bar
qq}(r_T,x)$, of interaction of a quark-antiquark dipole of
transverse separation $\vec r_T$ with a proton. The Bjorken
variable $x$ depends on $r_T$ and the dipole energy. Notice that
this is essentially a target rest frame description, interpreting
the beam hadron as a composition of different Fock states. Those
light-cone fluctuations are assumed to be "frozen" by Lorentz time
dilation during the interaction. Thus, the cross section is a sum
of the dipole cross sections for different Fock states. At the
same time, these dipole cross sections are averaged over the
properties of the target.

The paper is organized as follows. In Sect.~\ref{inclusive} we start with
inclusive production of heavy quarks and classify the different mechanisms. The
first one, which we call Bremsstrahlung, is similar (except for the couplings
and color factors) to the Drell-Yan mechanism. It arises from the interaction of
the radiating color charge, similar to bremsstrahlung in QED. The second one,
called Production mechanism, is related to direct interaction of the radiated
virtual gluon or the heavy quark pair with the target. The important observation
of this section is the smallness of the interference between the two mechanisms.

In Sect.~\ref{diff-qN} we calculate the forward amplitudes for
diffractive production of a $\bar QQ$ pair in quark-proton
collisions. The two mechanisms, bremsstrahlung and production, are
found to have different scale dependence. The former is a higher
twist effect, while the latter is leading twist and dominates the
diffractive cross section. We generalize these results to
diffractive gluon-proton collisions in Sect.~(\ref{gluons}).

The next step is the calculation of diffraction in proton-proton
collisions. In this case the Pomeron, which is a multigluon
exchange, can interact with active (radiating) and spectator
partons simultaneously. This possibility gives rise also to a
leading twist contribution for the bremsstrahlung mechanism.

In sect.~\ref{survival} we estimate the suppressing effects of
unitarity saturation, which is present in $pp$ collisions at high
energy. Our results are close to other estimates available in the
literature. The suppression caused by absorption ranges from an
order of magnitude at the Tevatron down to a few percent at LHC
energies.

Numerical calculations are performed in Sect.~\ref{results},
relying on the phenomenological dipole cross section, well fitted
to HERA data on the proton structure function. We present the
results for energy dependent diffractive cross sections of
different heavy flavor production, $x_1$ distribution (fractional
momentum carried by the heavy quark), and transverse momentum
dependence at different energies. We found that the bremsstrahlung
mechanism, although a leading twist, gives a negligibly small
contribution. The main part of the cross section corresponds to
the production mechanism in collisions of projectile quarks and
gluons with the target proton.

In Sect.~\ref{data} we review the available, rather scarce data on
diffractive production of heavy flavored hadrons. The data agree
quite well with our calculations. Our results and observations are
summarized in Sect.~\ref{summary}.

\section{Inclusive production at forward rapidities}\label{inclusive}

A comprehensive study of inclusive heavy flavor production within
the light-cone dipole approach was performed in \cite{kt-charm}
(see also review \cite{kr}). However, those calculations were
focused on heavy flavors produced at mid rapidities, where
gluon-gluon fusion is the dominant mechanism. Here we are
interested in heavy quarks produced in the projectile
fragmentation region, and our ultimate goal is diffractive
production. In this case the interaction with valence quarks must
be included, and this needs more elaborated calculations.

\begin{figure}[htbp]
 \includegraphics[width=4cm]{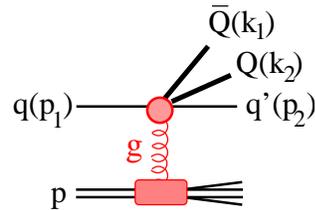}
 \caption{One gluon graph for heavy quark pair production by a quark in
inelastic collision}
 \label{graph}
 \end{figure}

First, we calculate the amplitude of inclusive cross section for
the production of a heavy pair $\bar QQ$ in a quark-proton
collision, $q+p\to q\bar QQX$, in one gluon approximation, as is
shown in Fig.~\ref{graph}.

In what follows we use the following notation (see
Fig.~\ref{graph}): $p_{1,2}$ are the 4-momenta of the projectile
quark q and the ejected quark $q^\prime$ respectively; $k_{1,2}$
are the 4-momenta of the produced heavy quarks $\bar Q$ and $Q$
respectively. For momenta combinations: $k=k_1+k_2$;
$\alpha=k^+/p_1^+$; $\beta=k^+_1/k^+$; $M^2=(k_1+k_2)^2$;
$Q^2=-(p_1-p_2)^2>0$; $\vec\pi=\alpha\vec p_2-(1-\alpha)\vec k$ is
the relative transverse momentum between $q^\prime$ and $(\bar
QQ)$; $\vec\kappa=(1-\beta)\vec k_1-\beta\vec k_2$ is the relative
transverse momentum between the heavy quarks.

To make further progress we switch from transverse momenta to
impact parameters, which require further definitions: $\vec r,\
\vec r_1,\ \vec r_2$ are the transverse separations within the
$\bar Q-Q$, $q-\bar Q$ and $q-Q$ pairs respectively;
$\vec\rho=\vec r-\beta\vec r_1-(1-\beta)\vec r_2$ is the distance
between $q'$ and the center of gravity of the $\bar QQ$ pair;
$\vec s=\vec r_1-\vec r_2$ is the $\bar QQ$ transverse separation.

\begin{figure}[htbp]
 \includegraphics[width=8cm]{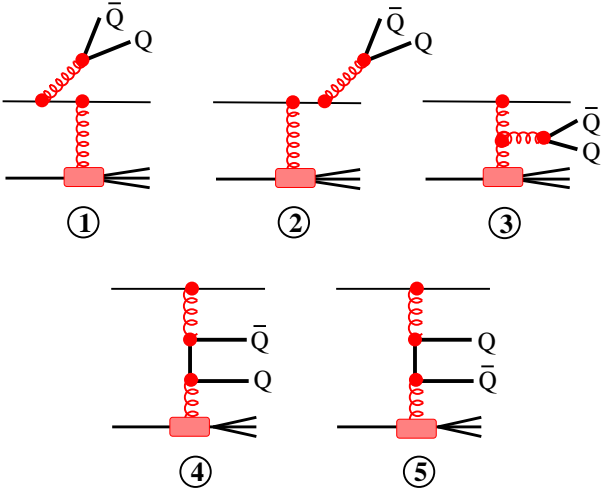}
 \caption{Feynman graphs contributing to inclusive production of a heavy
quark pair.}
 \label{graphs}
 \end{figure}

Since we are interested in $\bar QQ$ production in the projectile
fragmentation region, the rapidity interval between the $q'$ and
$\bar QQ$ is assumed to be short. At the same time the rapidity
interval between the $\bar QQ$ and the target is long (at high
energies) and is filled by radiated gluons. The bottom blob in the
graph in Fig.~\ref{graph} includes all those gluons, while the
upper part of the graph can be calculated using the Born
approximation which is rather accurate for the projectile
fragmentation region.  This part is represented by the Born graphs
depicted in Fig.~\ref{graphs}.
\vspace*{0cm}

The sum of this five amplitudes can be splited into two classes
which we assign to bremsstrahlung (Br) and production (Pr)
mechanisms, as is described in detail in \ref{classified},
 \beq
\sum\limits_{i=1}^5 M^{T(L)}_i=M^{T(L)}_{Br} + M^{T(L)}_{Pr}\ ,
\label{7}
 \eeq
 where the upper indexes $T$ and $L$ correspond to transverse or
longitudinal polarization of the virtual gluon radiated by the
projectile quark.

The transverse bremsstrahlung amplitude, $M^T_{Br}$, corresponds
to the following combination of graphs depicted in
Fig.~\ref{graphs} (see \ref{classified}.1),
 \beq
M^T_{Br}=M^T_1+M^T_2+\frac{Q^2}{M^2+Q^2}\,M^T_3\ . \label{10}
 \eeq
 It describes bremsstrahlung of a transversely polarized heavy gluon which
dissociates into $\bar QQ$. The remaining transverse amplitudes
are combined into the second group, the amplitude of $\bar QQ$
production via direct interaction with the heavy quark pair,
 \beq
M^T_{Pr} = \frac{M^2}{M^2+Q^2}\,M^T_3+M^T_4+M^T_5\ . \label{20}
 \eeq

The procedure of grouping the longitudinal amplitudes is more
complicated and is described in \ref{classified}.2. It leads also
to the structure Eq.~(\ref{7}), where the bremsstrahlung and
production longitudinal amplitudes are defined in (\ref{1.220}).

Thus, the inclusive cross section has the following structure,
 \beq
\frac{d\sigma(qp\to\bar QQX)}{d\alpha\,d\beta}=
\frac{d\sigma^T}{d\alpha\,d\beta}+
\frac{d\sigma^L}{d\alpha\,d\beta}\ , \label{24}
 \eeq
 where
 \beq
\frac{d\sigma^{T(L)}(qp\to\bar QQX)}{d\alpha\,d\beta}=
\frac{d\sigma^{T(L)}_{Br}}{d\alpha\,d\beta}+
\frac{d\sigma^{T(L)}_{Pr}}{d\alpha\,d\beta}+
\frac{d\sigma^{T(L)}_{Int}}{d\alpha\,d\beta}\ . \label{30}
 \eeq
 The last term here corresponds to the interference of the amplitudes
$M_{Br}$ and $M_{Pr}$.

\subsection{Bremsstrahlung mechanism}

The first term in the cross section Eq.~(\ref{30}) can be
calculated in the framework of the light-cone dipole approach as
follows,
 \beq
\frac{d\sigma_{Br}(qp\to\bar QQX)}{d\alpha\,d\beta} = \int
\frac{d^2\kappa}{(2\pi)^2}\,d^2\rho
\left|\Phi_{Br}(\vec\rho,\vec\kappa)\right|^2 \Sigma_1(\vec\rho);
\label{40}
 \eeq

Since the effective dipole cross section $\Sigma_1(\vec\rho)$ is a
function of one variable, for the sake of convenience we use here
the light-cone distribution amplitudes in mixed impact parameter -
transverse momentum representation,
 \beq
\Phi_{Br}(\vec\rho,\vec\kappa) = \sum\limits_\lambda
\Psi_1^{\lambda}(\vec\rho) \Psi_2^{\lambda}(\vec\kappa) +
\Psi_1^{L}(\vec\rho) \Psi_2^{L}(\vec\kappa)\ , \label{50}
 \eeq
 where the sum is over the transverse polarizations $\lambda$ of the
radiated virtual gluon; the amplitude of longitudinal gluon
radiation is labeled by the index $L$. Further notation is,
 \beqn
\Psi_1^{\lambda}(\vec\rho) &=&
\frac{\sqrt{\alpha_s(1/\rho)}}{2\pi}\, \chi_{f}^\dagger\left\{
i(2-\alpha)\vec e_{\lambda}\cdot\vec\nabla +
\alpha\left[\vec\sigma\times\vec e_{\lambda}\right]
\cdot\vec\nabla\right. \nonumber\\ &+& \left.
i\alpha^2m_q\left[\vec\sigma\times\vec e_{\lambda}\right]
\cdot\vec n\right\}\chi_{in}\, K_0(\tau\rho)\ ; \label{60}
 \eeqn
 \beqn
\Psi_1^{L}(\vec\rho)= \frac{\sqrt{\alpha_s(1/\rho)}}{2\pi}\,
\chi_{f}^\dagger\, 2(1-\alpha)M\,\chi_{in}\, K_0(\tau\rho)\ ,
\label{70}
 \eeqn
 which are the light-cone distribution amplitudes for the quark-gluon
Fock state with transversely polarized gluon (polarization
$\lambda$) and longitudinal gluon respectively. The spinors
$\chi_{in}$ and $\chi_{f}$ correspond to the initial and final
light quarks. The running QCD coupling $\alpha_s(1/\rho)$ is taken
at virtuality $\sim1/\rho^2$.

The distribution amplitudes for a $\bar QQ$ fluctuations of
transversely and longitudinally polarized gluons respectively,
read,
 \beqn
\Psi_2^{\lambda}(\vec\kappa) &=& \phi^\dagger\Bigl\{
m_q\vec\sigma\cdot\vec e_{\lambda} +
(1-2\beta)(\vec\sigma\cdot\vec n) (\vec
e_{\lambda}\cdot\vec\kappa) \nonumber\\ &+& i\left[\vec
e_{\lambda}\times\vec n\right] \cdot\vec\kappa\Bigr\}\,\bar\phi\,
\frac{\alpha_s(\kappa)}{m_Q^2+\kappa^2}\ ; \label{80}
 \eeqn
 \beqn
\Psi_2^{L}(\vec\kappa)= \phi^\dagger\,
2\beta(1-\beta)M\vec\sigma\cdot\vec n\, \bar\phi\,
\frac{\alpha_s(\kappa)}{m_Q^2+\kappa^2}\ , \label{90}
 \eeqn
 where
$\phi$, $\bar\phi$ are the spinors of the heavy $Q$ and $\bar Q$
respectively;
 \beqn
\tau^2 &=& (1-\alpha)M^2+\alpha^2m_q^2\ ; \nonumber\\
M^2&=&\frac{m_Q^2+\kappa^2}{\beta(1-\beta)}\ ; \nonumber
 \eeqn
 $m_q$ and $m_Q$ are the light and heavy quark masses respectively;
$\vec e_{\lambda}$ is the polarization vector of the transverse
gluon;  $\vec n$ is the unit vector aligned along the direction of
the projectile quark; $\vec\sigma$ are the Pauli matrices.

The dipole cross section $\Sigma_1(\rho)$ in (\ref{40})
corresponds to gluon radiation by a quark \cite{hir,kst1,kst2}. It
has the form of the cross section for a gluon-quark-antiquark
colorless system with transverse separations $\vec\rho$ for
gluon-quark, $(1-\alpha)\vec\rho$ for gluon-antiquark, and $\alpha
\vec\rho$ for quark-antiquark,
 \beq
\Sigma_1(\vec\rho)={9\over8}\left\{ \sigma(\vec\rho)+
\sigma\left[(1-\alpha)\vec\rho\right]\right\}
-{1\over8}\,\sigma(\alpha\vec\rho)\ , \label{100}
 \eeq
 where $\sigma(\rho)$ is the cross section of interaction of a $\bar qq$
dipole with transverse separation $\vec\rho$ on a proton.

\subsection{Production mechanism}

The second term in (\ref{30}) is represented in a form similar to
(\ref{40}),
 \beq
\frac{d\sigma_{Pr}(qp\to\bar QQX)}{d\alpha\,d\beta} = \int
\frac{d^2p_2}{(2\pi)^2}\,d^2 s \left|\Phi_{Pr}(\vec s,\vec
p_2)\right|^2 \Sigma_2(\vec s), \label{110}
 \eeq
 with  distribution amplitudes having structures similar to
(\ref{40})-(\ref{90}),
 \beq
\Phi_{Pr}(\vec s,\vec p_2) = \sum\limits_{\bar\lambda}
\Psi_3^{\bar\lambda}(\vec s) \Psi_4^{\bar\lambda}(\vec p_2) +
\Psi_3^{L}(\vec s) \Psi_4^{L}(\vec p_2)\ , \label{112}
 \eeq
 where
 \beqn
\Psi_3^{\bar\lambda}(\vec s) &=& \phi^\dagger\left\{
m_Q\vec\sigma\cdot\vec e_{\bar\lambda}
-
i(1-2\beta)(\vec\sigma\cdot\vec n) (\vec
e_{\bar\lambda}\cdot\vec\nabla) \right. \nonumber\\ &+& \left.
\left[\vec e_{\bar\lambda}\times\vec n\right]
\cdot\vec\nabla\right\}\,\bar\phi\, K_0(\epsilon s)\ ; \label{120}
 \eeqn
 \beqn
\Psi_3^{L}(\vec s)= \phi^\dagger\, 2\beta(1-\beta)Q\,\bar\phi\,
K_0(\epsilon s)\ , \label{130}
 \eeqn
 are the light-cone $\bar QQ$ distribution amplitudes in impact parameter
representation.  In the above expressions,
 \beqn
\epsilon^2&=&\beta(1-\beta)Q^2+m_Q^2\nonumber\ ;\\
Q^2&=&\frac{\vec p_2^{\,2}+\alpha^2 m_q^2}{1-\alpha}\ .
\label{135}
 \eeqn
 Notice that differently from the bremsstrahlung amplitudes,
here the gluon virtuality $Q$ is not equal to the effective mass
of the $\bar QQ$.

Correspondingly, the transverse and longitudinal distribution
amplitudes for quark-gluon fluctuation in momentum representation
read,
 \beqn
\Psi_4^{\bar\lambda}(\vec\kappa) &=&
\chi_{f}^\dagger\left\{(2-\alpha)\, \left(\vec
e_{\bar\lambda}\cdot\vec p_2\right) +
i\alpha\left[\vec\sigma\times\vec e_{\bar\lambda}\right] \cdot\vec
p_2\right. \nonumber\\ &+& \left. i\alpha^2
m_2\left[\vec\sigma\times\vec e_{\bar\lambda}\right] \cdot\vec
n\right\}\,\bar\chi_{in}\, \frac{1}{\alpha^2 m_q^2+p_2^2}\ ;
\label{140}
 \eeqn
 \beq
\Psi_4^{L}(\vec p_2)= 2(1-\alpha)\, \frac{\chi_{f}^\dagger\,
Q\chi_{in}}{\alpha^2 m_q^2+p_2^2}\ . \label{150}
 \eeq

The dipole cross section $\Sigma_2(\vec s)$ in (\ref{110})
corresponds to gluon decay into a quark-antiquark pair
\cite{hir,kt-charm}. Similar to (\ref{100}) it also has the form
of a cross section for a gluon-quark-antiquark colorless system,
but with transverse separations $\vec s$ for the quark-antiquark,
$\beta \vec\rho$ and $(1-\beta)\vec\rho$ for gluon-quark and
gluon-antiquark, respectively.
 \beq
\Sigma_2(\vec s)= {9\over8}\left\{ \sigma(\beta\vec s)+
\sigma\left[(1-\beta)\vec s\right]\right\} -
{1\over8}\,\sigma(\vec s)\ . \label{160}
 \eeq

\subsection{Bremsstrahlung-Production interference}

The third interference term in (\ref{30}) reads,
 \beqn
&&\frac{d\sigma_{Int}(qp\to\bar QQX)} {d\alpha\,d\beta} = 2{\rm
Re}\int d^2\rho\, d^2 s \nonumber\\ &\times& \left\la
\Phi^\dagger_{Pr}(\vec\rho,\vec s)\, \Phi_{Br}(\vec\rho,\vec
s)\right\ra\, \Sigma_{Int}(\vec\rho,\vec s). \label{175}
 \eeqn
 In this case the effective dipole cross section is function of two
coordinates,
 \beqn
\Sigma_{Int}(\vec\rho,\vec s) &=&
{7\over8}\Bigl\{\tilde\sigma(\vec\rho,-\beta\vec s)+
\tilde\sigma\bigl[(1-\alpha)\vec\rho,(1-\beta)\vec s\bigr]
\nonumber\\ &-& \tilde\sigma\bigl[\vec\rho,(1-\beta)\vec
s\bigr]\Bigr\} -
{1\over4}\Bigl\{\tilde\sigma\bigl[\vec\rho,(1-\beta)\vec s\bigr]
\nonumber\\ &+& \tilde\sigma\bigl[(1-\alpha)\vec\rho,-\beta\vec
s\bigr] - \tilde\sigma(\vec\rho,-\beta\vec s)\Bigr\}\ ,
\label{177}
 \eeqn
 where
 \beq
\tilde\sigma(\vec\rho,\vec s) = \sigma(\vec\rho+\vec
s)-\sigma(\vec\rho)- \sigma(\vec s)\ . \label{179}
 \eeq
 For this reason we use in (\ref{175}) the distribution functions $\Phi$
fully in impact parameter representation. They are related to the
mixed representation functions given in Eqs.~(\ref{50}) and
(\ref{112}) as,
 \beqn
\Phi^{T(L)}_{Br}(\vec\rho,\vec s) &=& \frac{1}{(2\pi)^2}\int
d^2\kappa\, e^{-i\vec\kappa\cdot\vec s}\,
\Phi^{T(L)}_{Br}(\vec\rho,\vec\kappa)\ ; \label{180}\\
\Phi^{T(L)}_{Pr}(\vec\rho,\vec s) &=& \frac{1}{(2\pi)^2}\int
d^2p_2\, e^{-i\vec p_2\cdot\vec s}\,
\Phi^{T(L)}_{Pr}(\vec\rho,\vec p_2)\ ; \label{182}
 \eeqn

This interference term, Eq.~(\ref{175}), turns out to be
suppressed in the cross section compared to the first two terms in
(\ref{30}). Indeed, performing the integrations in Eqs.~(\ref{40})
and (\ref{110}), one arrives at cross sections which expose the
leading twist scale dependence both for the bremsstrahlung and
production mechanisms,
 \beq
\frac{d\sigma_{Br}}{d\alpha\,d\beta} \sim
\frac{d\sigma_{Pr}}{d\alpha\,d\beta} \sim \frac{1}{m_Q^2}\ .
\label{186}
 \eeq
 Indeed, the mean transverse separations squared, which are controlled by
the distribution amplitudes Eqs.~(\ref{60}), (\ref{70}) and
Eqs.~(\ref{120}), (\ref{130}), are $\la\rho^2\ra\sim
1/\epsilon^2\sim 1/m_Q^2$ and $\la s^2\ra\sim 1/\tau^2\sim
1/m_Q^2$, respectively. At such small separations the cross
sections Eqs.~(\ref{100}) and (\ref{160}) behave in accordance
with color transparency \cite{zkl},
$\Sigma_1(\vec\rho)\propto\rho^2$ and $\Sigma_2(\vec s)\propto
s^2$, confirming (\ref{186}). This result could be expected, since
it also follows from the QCD factorization scheme.

As for the interference term, it follows from Eqs.~(\ref{50}) and
(\ref{112}) that the product of distribution amplitudes in
(\ref{175}), $\left\la \Phi^\dagger_{Pr}(\vec\rho,\vec
s)\,\Phi_{Br}(\vec\rho,\vec s)\right\ra$, contains only
coordinates squared, $\vec\rho^{\,2}$ and $\vec s^{\,2}$, but not
their product, $\vec\rho\cdot\vec s$. At the same time, since the
dipole cross section at small separation behaves as $\sigma(r)=
C\,r^2$, the combination of cross sections, Eq.~(\ref{179}),
reads,
 \beq
\tilde\sigma(\vec\rho,\vec s)=2C\,\vec\rho\cdot\vec s\ .
\label{188}
 \eeq
 Thus, the interference term Eq.~(\ref{135}) vanishes after integration
over azimuthal angle.

In a more realistic model with a dipole cross section leveling off at large separations,
Eq.~(\ref{450}). the result is not zero, but is a higher twist effect,
 \beq
\frac{d\sigma_{Int}}{d\alpha\,d\beta} \sim
\frac{1}{m_Q^4R^2_0(x)}\ , \label{190}
 \eeq
 where $R_0(x)$ is defined in (\ref{450}). Thus, the interference term in
the cross section of inclusive heavy flavor production is
vanishingly small. This observation turns out to be valid for
diffractive production as well. Therefore, we safely neglect the
interference in what follows.

\section{Diffractive production of heavy flavors in quark-proton
collisions}\label{diff-qN}

\begin{figure}[htbp]
 \includegraphics[width=4cm]{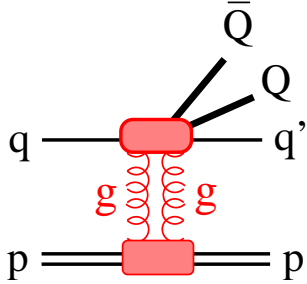}
 \caption{Diffractive production of a heavy quark pair in a quark-proton
collision}
 \label{graph-diff}
 \end{figure}

Here we calculate the cross section of diffractive excitation of a
projectile quark resulting in the production of a heavy quark
pair,
 \beq
q+p\to q\,\bar QQ+p\ , \label{200}
 \eeq
 as is shown in Fig.~\ref{graph-diff}.
 This picture represents numerous Feynman graphs which are
equivalent to a rather simple factorized form of the diffractive
amplitude in the light-cone dipole representation. This formalism
was developed for diffractive gluon bremsstrahlung in \cite{kst2}.

Similar to (\ref{10})-(\ref{20}), the diffractive amplitude
$A^{qp}$ can be splited into two parts,
 \beq
A^{qp}=A_{Br}^{qp}+A_{Pr}^{qp}\ , \label{210}
 \eeq
 where the bremsstrahlung and production amplitudes are defined as follows.

\subsection{Bremsstrahlung mechanism of diffraction}

The bremsstrahlung amplitude in (\ref{210}) has the form,
 \beq
A^{qp}_{Br}(\vec\pi,\vec\kappa) = \frac{3i}{16\pi} \int
d^2\rho\,d^2s\, e^{i\vec\pi\cdot\vec\rho+ i\vec\kappa\cdot\vec
s}\, \Phi_{Br}(\vec\rho,\vec s)\, \tilde\Sigma_1(\vec\rho,\vec s)\
. \label{220}
 \eeq
 The distribution amplitude $\Phi_{Br}(\vec\rho,\vec s)$ in coordinate
space is given by Eq.~\ref{180}),
 \beqn
\Phi_{Br}(\vec\rho,\vec s) &=& \Phi^L_{Br}(\vec\rho,\vec s) +
\Phi^T_{Br}(\vec\rho,\vec s)=
\frac{\sqrt{\alpha_s(1/s)\alpha_s(1/\rho)}} {(2\pi)^2} \nonumber\\
&\times& \Bigl[4(1-\alpha)\frac{m_Q}{x} K_1(m_Qx)\chi_f\chi_{in}
\phi^\dagger\vec\sigma\vec n\phi \nonumber\\ &+&
\sum\limits_\lambda a_\lambda(\vec\rho)\,
b_\lambda(\vec\nabla_s)\, \frac{1}{\rho^2}\,K_0(m_Qx)\Bigr]\ ,
\label{230}
 \eeqn
 where
\beq a_\lambda(\vec\rho) = \chi_f\Bigl\{ i(2-\alpha)\vec
e_\lambda\cdot\vec\rho + \alpha[\vec\sigma\times\vec e_\lambda]
\cdot\vec\rho\Bigr\}\chi_{in}; \label{232}
 \eeq
 \beqn
b_\lambda(\vec\nabla_s) &=& \phi\Bigl\{ m_q\vec\sigma\cdot\vec
e_\lambda + [\vec e_\lambda\times\vec n] \cdot\vec\nabla_s
\nonumber\\&-& i(1-2\beta)(\vec\sigma\cdot n) (\vec
e_\lambda\cdot\nabla_s) \Bigr\}\bar\phi; \label{233}\\ x&=&
\sqrt{s^2+\frac{(1-\alpha)\rho^2} {\beta(1-\beta)}}\ . \label{234}
 \eeqn

\begin{figure}[htbp]
 \includegraphics[width=8cm]{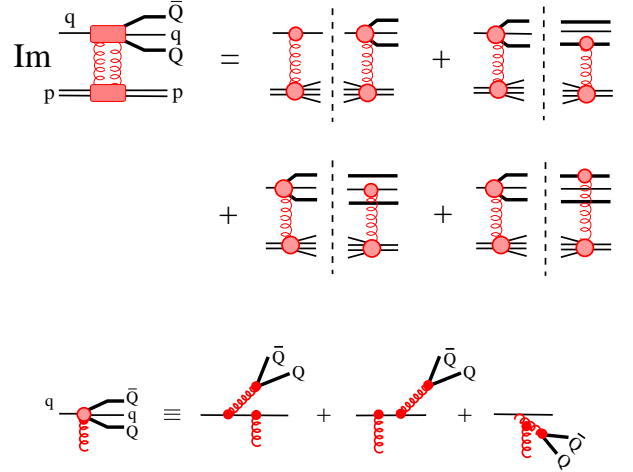}
 \caption{Cutkosky rules for the imaginary part of the amplitude
of diffractive $\bar QQ$ production in quark-proton collision.
Only bremsstrahlung mechanism is considered for this example.}
 \label{cutk}
 \end{figure}

 The imaginary part of the amplitude of diffractive production of heavy
flavors is calculated employing the generalized optical theorem
(Cutkosky cutting rules) \cite{cutkosky}. It is illustrated in
Fig.~\ref{cutk} for the example of the bremsstrahlung mechanism,
and is used in the same way for other mechanisms in what follows.
 The amplitudes shown in the graphs on both sides of the unitarity cut
(vertical dashed line) are on-mass-shell physical processes, and
the final state excitations of the proton are summed up employing
completeness. After integration over impact parameter $\vec b$,
this procedure leads to the following form for the effective cross
section $\tilde\Sigma_1$ in (\ref{220}),
 \beqn
&&\tilde\Sigma_1(\vec\rho,\vec s)= \left({1\over9}+{5\over24}\,
\tau^q_a\tau^Q_a\right) \Bigl\{\sigma\left[\vec\rho-(1-\beta)\vec
s\right] \nonumber\\&-& \sigma\left[\vec\rho+\beta\vec s\right] -
\sigma\left[(1-\alpha)\vec\rho-(1-\beta)\vec s\right]
\nonumber\\&+& \sigma\left[(1-\alpha)\vec\rho+\beta\vec s\right]
\Bigr\} + {3\over4}\,\tau^q_a \tau^Q_a\Bigl\{ \sigma[\vec\rho] -
\sigma[(1-\alpha)\vec\rho] \nonumber\\&+& \sigma[(1-\beta)\vec s]
- {1\over2}\, \sigma[(1-\alpha)\vec\rho-(1-\beta)\vec s]
\nonumber\\&-& {1\over2}\,\sigma[\vec\rho-(1-\beta)\vec s] -
\sigma[\beta\vec s] \nonumber\\&+&
{1\over2}\,\sigma[(1-\alpha)\vec\rho-\beta\vec s] +
{1\over2}\,\sigma[\vec\rho+\beta\vec s]\Bigr\}\ . \label{240}
 \eeqn
 Here $\tau^q_a={1\over2}\lambda^q_a$ and
$\tau^Q_a={1\over2}\lambda^Q_a$, where $\lambda^q_a$ and
$\lambda^Q_a$ are the Gell-Mann matrices acting on the color
spaces of light and heavy quarks respectively.

The corresponding bremsstrahlung term in the diffractive cross
section reads,
 \beqn
&&\left.\frac{d\sigma_{Br}(qp\to\bar QQqp)} {dt'\,d\alpha\,
d\beta}\right|_{t'=0} = \frac{3}{256\pi} \int d^2\rho\,d^2 s
\nonumber\\ &\times& \left|\Phi_{Br}(\vec\rho,\vec s)\right|^2\,
{\rm Tr}_q{\rm Tr}_Q\left[\tilde\Sigma_1(\vec\rho,\vec s)
\tilde\Sigma_1^\dagger(\vec\rho,\vec s)\right]\ , \label{250}
 \eeqn
 where the traces Tr$_q$ and Tr$_Q$ are taken over the Gell-Mann
matrices corresponding to light and heavy quarks respectively.

\subsection{Production mechanism of diffraction}\label{q-pr}

The production term in (\ref{210}) reads,
 \beq
A^{qp}_{Pr} = \frac{3i}{16\pi} \int d^2\rho\,d^2s\,
e^{i\vec\pi\cdot\vec\rho+ i\vec\kappa\cdot\vec s}\,
\Phi_{Pr}(\vec\rho,\vec s)\, \tilde\Sigma_2(\vec\rho,\vec s)\ ,
\label{260}
 \eeq
 where
 \beqn
&&\Phi_{Pr}(\vec\rho,\vec s) = \Phi^L_{Pr}(\vec\rho,\vec s) +
\Phi^T_{Pr}(\vec\rho,\vec s) =
\frac{\sqrt{\alpha_s(1/s)\alpha_s(1/\rho)}} {(2\pi)^2} \nonumber\\
&\times& \Bigl\{4(1-\alpha)\frac{m_Q}{x} K_1(m_Qx)\chi_f\chi_{in}
\phi^\dagger\vec\sigma\vec n\phi \nonumber\\ &+&
\sum\limits_\lambda a_\lambda(\vec\rho)\,
b_\lambda(\vec\nabla_s)\,
\frac{1}{\rho^2}\bigl[K_0(m_Qs)-K_0(m_Qx)\Bigr\}\ , \label{270}
 \eeqn
 The functions $a_\lambda(\vec\rho)$, $b_\lambda(\vec\nabla_s)$ and $x$
are defined in (\ref{232})-(\ref{234}).

 \beqn
&&\tilde\Sigma_2(\vec\rho,\vec s)= \left({1\over9}+{5\over24}\,
\tau^q_a\tau^Q_a\right) \Bigl\{\sigma\left[\vec\rho+\beta\vec
s\right] \nonumber\\&-& \sigma\left[\vec\rho-(1-\beta)\vec
s\right] + \sigma\left[(1-\alpha)\vec\rho+\beta\vec s\right]
\nonumber\\&-& \sigma\left[(1-\alpha)\vec\rho-(1-\beta)\vec
s\right] \Bigr\} + {7\over12}\,\tau^q_a \tau^Q_a\sigma[\vec s]+
\nonumber\\&+& {3\over4}\,\tau^q_a \tau^Q_a\Bigl\{ \sigma[\vec s]
- \sigma[(1-\beta)\vec s] -\sigma[\beta\vec s] + \sigma(\vec\rho)
\nonumber\\&-& \sigma[(1-\alpha)\vec\rho] +
 {1\over2}\,
\sigma[(1-\alpha)\vec\rho-(1-\beta)\vec s] \nonumber\\&+&
{1\over2}\,\sigma[(1-\alpha)\vec\rho+\beta\vec s] -
{1\over2}\,\sigma[\vec\rho-(1-\beta)\vec s] \nonumber\\&-&
{1\over2}\,\sigma[\vec\rho+\beta\vec s]\Bigr\}\ . \label{280}
 \eeqn

Eventually we arrive at the production cross section,
 \beqn
&&\left.\frac{d\sigma_{Pr}(qp\to\bar QQqp)} {dt'\,d\alpha\,
d\beta}\right|_{t'=0} = \frac{3}{256\pi} \int d^2\rho\,d^2 s
\nonumber\\ &\times& \left|\Phi_{Pr}(\vec\rho,\vec s)\right|^2\,
{\rm Tr}_q{\rm Tr}_Q \left[\tilde\Sigma_2(\vec\rho,\vec s)
\tilde\Sigma_2^\dagger(\vec\rho,\vec s)\right]\ . \label{290}
 \eeqn

\subsection{Scale dependence}

In the bremsstrahlung distribution amplitude, Eq.~(\ref{230}),
both mean separations are controlled by the hard scale,
 \beq
\la\rho^2\ra \sim \la s^2\ra \sim \frac{1}{m_Q^2}\ . \label{300}
 \eeq
 Therefore, the corresponding term Eq.~(\ref{250}) in the diffractive
cross section is as small as $1/m_Q^4$, i.e. it is a higher twist
effect.

On the contrary, in the production mechanism only the $\bar QQ$
separation is small, $\la s^2\ra\sim 1/m_Q^2$. The mean separation
between the light quark $q'$ and the $\bar QQ$ according to
(\ref{140})-(\ref{150}) is large, $\la\rho^2\ra\sim 1/m_q^2$.
However, the effective cross section Eq.~(\ref{280}) cannot be
large. Indeed, at small $s\to 0$ it vanishes as
 \beq
\tilde\Sigma_2(\vec\rho,\vec s) \propto \vec s\cdot\vec\rho\ .
\label{310}
 \eeq
 This result is similar to what was found in \cite{kpst-dy} for the diffractive
Drell-Yan reaction, which also probes simultaneously large and
small distances. This is very nontrivial, since in the case of the
Drell-Yan reaction that property is due to the Abelian nature of
the radiated particle. Now we have a non-Abelian radiation, but
arrived at the same feature.

It is interesting to notice that while the forward Abelian
radiation by a quark is forbidden, the bremsstrahlung part of the
diffractive radiation of a $\bar QQ$ pair, although is not zero,
turns out to be quite suppressed.

For further calculations we can safely neglect the higher twist
contribution of the bremsstrahlung mechanism to the quark-proton
diffractive amplitude, and keep only the leading twist production
term, Eq.~(\ref{290}).

\section{Diffractive production of heavy flavors by a
gluon}\label{gluons}

\begin{figure}[htbp]
 \includegraphics[width=4cm]{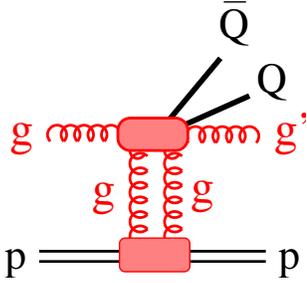}
 \caption{Diffractive production of a heavy quark pair in a gluon-proton
collision}
 \label{glue-diff}
 \end{figure}

Not only quarks, but projectile gluons can also be diffractively
excited producing a heavy $\bar QQ$ pair, as is illustrated in
Fig.~\ref{glue-diff}.
 This contribution is important if the $\bar QQ$ pair carries a small
fraction of the initial beam momentum.

The calculations are similar to those that we performed in
Sect.~\ref{q-pr}. Here we neglect the higher twist bremsstrahlung
contribution. The leading twist amplitude of diffractive $\bar QQ$
pair production by a projectile gluon, $g+p\to \bar QQX+p$, reads,
 \beq
A^{(Gp)}_{Pr}=\frac{3i}{16\pi} \int d^2\rho\,d^2s\,
e^{i\vec\pi\cdot\vec\rho + i\vec\kappa\cdot\vec s}\,
\Phi^{(Gp)}_{Pr}(\vec\rho,\vec s)\, \tilde\Sigma_3(\vec\rho,\vec
s)\ , \label{312}
 \eeq
 where
 \beqn
\Phi^{(Gp)}_{Pr}(\vec\rho,\vec s) \hspace{-1mm}&=&
\frac{\sqrt{\alpha_s(1/\rho)\alpha_s(1/s)}} {(2\pi)^2}\,
\Bigl\{4\frac{m_Q}{x}\, K_1(m_Qx)\,\vec e_f\cdot\vec e_{in}
\nonumber\\ &+& \hspace{-2mm} \sum\limits_\lambda
a_\lambda(\vec\rho) b_\lambda(\vec\nabla_s)
\frac{1}{\rho^2}\bigl[K_0(m_Q s)-K_0(m_Q x)\bigr]
\Bigr\}\nonumber\\ \label{314}
 \eeqn
 Here
 \beqn
a_\lambda(\vec\rho) &=& 2i\bigl[(1-\alpha) (\vec e_f\cdot\vec
e_{in}) (\vec e_\lambda\cdot\vec\rho) + \alpha(\vec
e_{in}\cdot\vec e_\lambda) (\vec e_f\cdot\vec\rho) \nonumber\\ &-&
\alpha(\vec e_f\cdot\vec e_\lambda) (\vec e_i\cdot\vec\rho)\bigr]\
, \label{315}
 \eeqn
 $b_\lambda(\vec\nabla_s)$ is given by Eq.~(\ref{233}), $x$ is given
by (\ref{234}), $\vec e_{in}$ and $\vec e_f$ are the polarization
vectors of gluons in the initial and final states respectively,
and $\vec e_\lambda$ is the transverse polarization vector of the
virtual $t$-channel gluon. The first and the second terms in the
r.h.s. of Eq.~(\ref{314}) correspond to exchange in the
$t$-channel of longitudinally and transversely polarized gluons
respectively.

 The effective cross section in (\ref{312}) is given by
 \beqn
\tilde\Sigma_3(\vec\rho,\vec s) &=& \left({1\over4}\delta_{ab} +
{3\over8}d_{abc}\tau_c\right)
\Bigl\{\sigma\bigl[(1-\alpha)\vec\rho- (1-\beta)\vec s\bigr]
\nonumber\\ &+& \sigma\bigl[\vec\rho-(1-\beta)\vec s\bigr]-
\sigma\bigl[(1-\alpha)\vec\rho+ \beta\vec s\bigr] \nonumber\\ &-&
\sigma\bigl[\vec\rho+\beta\vec s\bigr] \Bigr\} +
\frac{3i}{8}\,f_{abd}\tau_d
\Bigl\{2\sigma\bigl[(1-\alpha)\rho\bigr] \nonumber\\ &-&
\sigma\bigl[(1-\alpha)\vec\rho+ \beta\vec s\bigr]-
\sigma\bigl[(1-\alpha)\vec\rho- (1-\beta)\vec s\bigr]
\nonumber\\&-& 2\sigma(\rho) + \sigma\bigl[\vec\rho+\beta\vec
s\bigr]+ \sigma\bigl[\vec\rho-(1-\beta)\vec s\bigr] \nonumber\\
&-& 2\sigma(\beta s) - 2\sigma\bigl[(1-\beta) s\bigr] +
{32\over9}\sigma(s)\Bigr\}\,. \label{316}
 \eeqn
 The indexes $a,\ b$ correspond to the initial and final state polarizations
of the gluons, and $d_{abc}$ and $f_{abc}$ are the structure
constants.

Eventually, the differential cross section of diffractive
production of a $\bar QQ$ pair in gluon-proton collision by the
production mechanism (leading twist) turns out to be
 \beqn
&& \left. \frac{d\sigma_{Pr}(Gp\to G\bar QQp)}
{dt'\,d\alpha\,d\beta}\right|_{t'=0} \nonumber\\ &=&
\frac{9}{256\pi} \int d^2\rho d^2s
\left|\Phi^{(Gp)}_{Pr}(\vec\rho,\vec s)\right|^2 {\rm Tr}_Q
\left\la\tilde\Sigma_3(\vec\rho,\vec s)
\tilde\Sigma_3^\dagger(\vec\rho,\vec s)\right\ra \nonumber\\
\label{317}
 \eeqn
 Here the brackets $\bigl\la ...\bigr\ra$ indicate the averaging over
initial and summing over final color indexes of the gluons.

\section{Diffractive proton-proton collisions}

Now we consider a large rapidity gap single diffractive process,
 \beq
p+p\to \bar QQX + p \,, \label{320}
 \eeq
 where one colliding proton remains intact, and the debris of the other
proton contains a heavy $\bar QQ$ pair. Similar to the
quark-proton collision, the amplitude of this reaction can be
splited into bremsstrahlung and production parts,
 \beq
A^{pp}=A^{pp}_{Br}+A^{pp}_{Pr}\ , \label{330}
 \eeq
 which are described below.

\subsection{Leading twist bremsstrahlung contribution}

\begin{figure}[htbp]
 \includegraphics[width=7cm]{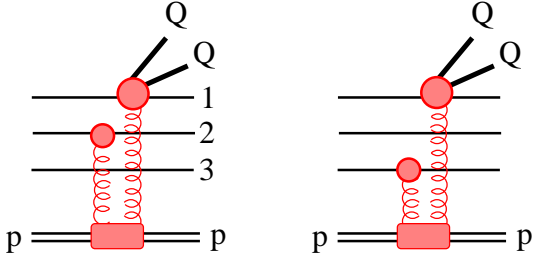}
 \caption{Diffractive production of a heavy quark pair in a proton-proton
collision. More graphs emerge after quark permutation.}
 \label{pp-br}
 \end{figure}

The first bremsstrahlung term in (\ref{330}) is represented by
graphs depicted in Fig.~\ref{pp-br}.
 We neglect the higher twist term corresponding to attachment
of both t-channel gluons to the same valence quark in the
projectile proton. More diagrams are generated by permutation of
the projectile quarks.

The diffractive amplitude corresponding to the graphs in
Fig.~\ref{pp-br} has the form,
 \beqn
A_{Br}^{pp}&=&\frac{3i}{16\pi} \int d^2\rho\,d^2s\,
\prod\limits_{i=1}^3 d^2r_i\,dx_i \, e^{i\vec\pi\cdot\vec\rho +
i\vec\kappa\cdot\vec s} \nonumber\\ &\times& \Psi_f^\dagger(\vec
r_1,\vec r_2,\vec r_3,x_1,x_2,x_3)\,
\Sigma^{pp}_{Br}(\vec\rho,\vec s,\vec r_1,\vec r_2,\vec r_3)
\nonumber\\ &\times& \Phi_{Br}(\vec\rho,\vec s)\,
\Psi_{in}^\dagger(\vec r_1,\vec r_2,\vec r_3,x_1,x_2,x_3).
\label{340}
 \eeqn
 Here $\Psi_{in}$ is the light-cone wave function of the valence $|3q\ra$
Fock component of the projectile proton, and $\Psi_{f}$ is the
final state wave function of the recoil $|3q\ra$ system after
radiation of the $\bar QQ$.

The effective dipole cross section has the following structure,
 \beqn
&&\Sigma_{Br}^{pp}(\vec\rho,\vec s,\vec r_1,\vec r_2,\vec r_3)=
\nonumber\\ && \left[{1\over6}\,\delta_{ab} +
{1\over2}\,d_{abc}\tau^{q_1}_c\right]\,\tau^Q_b
\left[\tau^{q_2}_a\sigma_1(\vec r_{12},\vec\rho)+
\tau^{q_3}_a\sigma_1(\vec r_{13},\vec\rho)\right] \nonumber\\ &+&
{i\over2}\,f_{abc}\tau^{q_1}_c \tau^Q_b
\left[\tau^{q_2}_a\sigma_2(\vec r_{12},\vec\rho)+
\tau^{q_3}_a\sigma_2(\vec r_{13},\vec\rho)\right] \nonumber\\ &+&
{\rm quark\ permutations}\ , \label{350}
 \eeqn
 where the upper index $q_i$ ($i=1,2,3$) indicates the active quarks.

 The following new notation is introduced here,
 \beqn
\sigma_1(\vec r,\vec\rho)&=& \sigma(\vec r+\alpha\vec\rho)-
\sigma(\vec r)\ , \nonumber\\ \sigma_2(\vec r,\vec\rho)&=&
2\sigma[\vec r-(1-\alpha)\vec\rho]- \sigma(\vec r+\alpha\vec\rho)-
\sigma(\vec r)\ , \nonumber\\ \vec r_{12}&=&\vec r_1-\vec r_2\ ,\
\ \ \ \vec r_{13}=\vec r_1-\vec r_3\ . \label{360}
 \eeqn

The differential cross section corresponding to the amplitude
Eq.~(\ref{340}) reads,
 \beq
\frac{d\sigma_{Br}(pp\to\bar QQXp)} {dt'\,dz_1\,dz_2} = \int
d^2\kappa\,d^2\pi\, \sum\limits_{f}\left|A_{Br}^{pp}\right|^2\ ,
\label{370}
 \eeq

One can employ completeness for the final state $|3q\ra$ system,,
\beqn &&\sum\limits_{f} \Psi_f(\vec r_1,\vec r_2,\vec
r_3,x_1,x_2,x_3) \Psi_f^\dagger(\vec r_1^{\,\prime},\vec
r_2^{\,\prime},\vec
r_3^{\,\prime},x_1^\prime,x_2^\prime,x_3^\prime) \nonumber\\& =&
\prod\limits_i^3 \delta(\vec r_i-\vec r_i^{\,\prime})\,
\delta(x_i-x_i^\prime)\ . \label{380}
 \eeqn
 Applying this condition to (\ref{370}) and averaging over color indexes
we get,
 \beqn
&&\frac{d\sigma_{Br}(pp\to\bar QQXp)} {dt'\,dz_1\,dz_2} =
\frac{9}{256\pi}\int d^2\rho\,d^2s \prod\limits_i^3 d^2r_i\,dx_i
\nonumber\\ &\times& \Bigl|\Psi_{in}(\vec r_1,\vec r_2,\vec
r_3,x_1,x_2,x_3)\Bigr|^2 \,\Bigl|\Phi_{Br}(\vec\rho,\vec
s)\Bigr|^2 \nonumber\\ &\times& {\rm Tr}_Q\left\la
\Sigma^{pp}_{Br}(\vec\rho,\vec s,\vec r_1,\vec r_2,\vec r_3)
\left.\Sigma^{pp}_{Br}\right.^\dagger (\vec\rho,\vec s,\vec
r_1,\vec r_2,\vec r_3) \right\ra\ . \nonumber\\ \label{390}
 \eeqn
 Here the trace is performed only over color indexes of the $\bar QQ$ pair, while
the brackets indicate averaging over the colorless $|3q\ra$ state
in the proton. This trace results in,
 \beqn
&&{\rm Tr}_Q\left\la \Sigma^{pp}_{Br}(\vec\rho,\vec s,\vec
r_1,\vec r_2,\vec r_3) \left.\Sigma^{pp}_{Br}\right.^\dagger
(\vec\rho,\vec s,\vec r_1,\vec r_2,\vec r_3) \right\ra \nonumber\\
&=& {109\over108}\,\sigma^2_1(\vec r_{12},\vec\rho) +
{1\over4}\,\sigma^2_2(\vec r_{12},\vec\rho) -
{1\over6}\,\sigma_1(\vec r_{12},\vec\rho)\sigma_2(\vec
r_{12},\vec\rho) \nonumber\\ &+& {11\over216}\,\sigma_1(\vec
r_{12},\vec\rho) \sigma_1(\vec r_{13},\vec\rho)+
{1\over8}\,\sigma_2(\vec r_{12},\vec\rho) \sigma_2(\vec
r_{13},\vec\rho) \nonumber\\ &-& {5\over14}\,\sigma_1(\vec
r_{12},\vec\rho) \sigma_2(\vec r_{13},\vec\rho) \nonumber\\ &+&
\{\vec r_{12} \rightleftharpoons \vec r_{13}\} + {\rm quark\
permutations}\ . \label{400}
 \eeqn

Further integrations in Eq.~(\ref{390}) are straightforward.

\subsection{Leading twist production mechanism}\label{feynman}

\begin{figure}[htbp]
 \includegraphics[width=8cm]{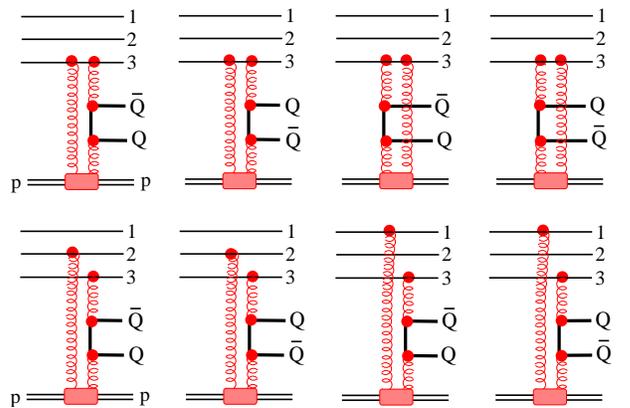}
 \caption{Diffractive production graphs providing $1/m_Q^2$ contribution
to the cross section. The projectile valence quarks should be
permuted.}
 \label{pp-pr}
 \end{figure}

In this case the main contribution emerges from the diffractive
interaction of a separate valence quark, although the interaction
of spectators should be included as well. The full set of graphs
having the leading twist behavior, $1/m_Q^2$, in the cross section
is shown in Fig.~\ref{pp-pr}).
 In all these graphs two gluons are attached to the target and beam
protons. The former is obvious, since we want to have a large
rapidity gap and also that the recoil target proton remains
intact. As for two gluons attached to the beam, this condition is
due to the leading twist behavior we would like to have. Indeed,
in the set of graph where only one gluon is attached to the
projectile, the second $t$-channel gluon is probing the small
sizes $r\sim 1/m_Q$ in the dissociation process $G\to\bar QQ$.
Thus one gains an additional suppression $1/m_Q^2$. i.e. a higher
twist behavior. Notice, however, that the pairs of gluons attached
to the target and the beam are not symmetric, the former are in a
colorless state, while in the latter all colors are summed up.

 The cross section corresponding to the graphs depicted in
Fig.~\ref{pp-pr} has a form similar to (\ref{390}),
 \beqn
&&\frac{d\sigma_{Pr}(pp\to\bar QQXp)} {dt'\,dz_1\,dz_2} =
\frac{9}{256\pi}\int d^2\rho\,d^2s \prod\limits_i^3 d^2r_i\,dx_i
\nonumber\\ &\times& \Bigl|\Psi_{in}(\vec r_1,\vec r_2,\vec
r_3,x_1,x_2,x_3)\Bigr|^2 \,\Bigl|\Phi_{Pr}(\vec\rho,\vec
s)\Bigr|^2 \nonumber\\ &\times& {\rm Tr}_Q\left\la
\Sigma^{pp}_{Pr}(\vec\rho,\vec s,\vec r_1,\vec r_2,\vec r_3)
\left.\Sigma^{pp}_{Pr}\right.^\dagger (\vec\rho,\vec s,\vec
r_1,\vec r_2,\vec r_3) \right\ra\ . \nonumber\\ \label{410}
 \eeqn
 In this case the effective dipole cross section has the form,
 \beqn
&&\Sigma^{pp}_{Pr}(\vec\rho,\vec s,\vec r_1,\vec r_2,\vec r_3)=
\nonumber\\ && (\tau^{q_1}_a\tau^{q_1}_b)\,\left[
(\tau^Q_a\tau^Q_b)\,\sigma_3(\vec\rho,\vec s) +
(\tau^Q_b\tau^Q_a)\,\sigma_4(\vec\rho,\vec s)\right] \nonumber\\
&+& \tau^{q_1}_a(\tau^Q_a\tau^Q_b)\Bigl[\tau^{q_2}_b\,
\sigma_5(\vec r_{12},\vec\rho,\vec s) +
\tau^{q_3}_b\,\sigma_5(\vec r_{13},\vec\rho,\vec s) \nonumber\\
&+& \tau^{q_1}_a(\tau^Q_b\tau^Q_a)\Bigl[\tau^{q_2}_b\,
\sigma_6(\vec r_{12},\vec\rho,\vec s) +
\tau^{q_3}_b\,\sigma_6(\vec r_{13},\vec\rho,\vec s)\Bigr]\ '
\nonumber\\ &+& {\rm quark\ permutations}\ . \label{420}
 \eeqn
 where
 \beqn
\sigma_3(\vec\rho,\vec s)&=& {1\over2}\Bigl\{\sigma(\vec\rho)-
\sigma(\vec\rho-\beta\vec s) - \sigma[(1-\alpha)\vec\rho]
\nonumber\\ &+& \sigma[(1-\alpha)\vec\rho+(1-\beta)\vec s]\;
\Bigr\} \nonumber\\ \sigma_4(\vec\rho,\vec s)&=&
{1\over2}\Bigl\{\sigma[\vec\rho+(1-\beta)\vec s]
-\sigma(\vec\rho)+ \sigma[(1-\alpha)\vec\rho] \nonumber\\ &-&
\sigma[(1-\alpha)\vec\rho-\beta\vec s]\Bigr\}\ ; \nonumber\\
\sigma_5(\vec r_{1i},\vec\rho,\vec s)&=& \sigma[\vec
r_{1i}-(1-\alpha)\vec\rho] \nonumber\\ &-& \sigma[\vec
r_{1i}-(1-\alpha)\vec\rho+\beta\vec s]\ ; \nonumber\\
\sigma_6(\vec r_{1i},\vec\rho,\vec s)&=& \sigma[\vec
r_{1i}-(1-\alpha)\vec\rho-(1-\beta)\vec s] \nonumber\\&-&
\sigma[\vec r_{1i}-(1-\alpha)\vec\rho]\ , \nonumber\\ i&=&2,\ 3\ .
\label{430}
 \eeqn

The trace of the product of the effective cross sections in
(\ref{410}) has the form,
 \beqn
&&{\rm Tr}_Q\left\la \Sigma^{pp}_{Pr}(\vec\rho,\vec s,\vec
r_1,\vec r_2,\vec r_3) \left.\Sigma^{pp}_{Pr}\right.^\dagger
(\vec\rho,\vec s,\vec r_1,\vec r_2,\vec r_3) \right\ra \nonumber\\
&=& {19\over6}\,\sigma_3^2(\vec\rho,\vec s) +
{2\over3}\,\sigma_4^2(\vec\rho,\vec s) -
{2\over3}\,\sigma_3(\vec\rho,\vec s)\, \sigma_4(\vec\rho,\vec s)
\nonumber\\ &+& {1\over9}\,\sigma_4(\vec\rho,\vec s)
\left[\sigma_5(\vec r_{12},\vec\rho,\vec s) + \sigma_5(\vec
r_{13},\vec\rho,\vec s)\right] \nonumber\\ &-&
{2\over9}\,\sigma_4(\vec\rho,\vec s) \left[\sigma_6(\vec
r_{12},\vec\rho,\vec s) + \sigma_6(\vec r_{13},\vec\rho,\vec
s)\right] \nonumber\\ &-& {19\over18}\,\sigma_3(\vec\rho,\vec s)
\left[\sigma_5(\vec r_{12},\vec\rho,\vec s) + \sigma_5(\vec
r_{13},\vec\rho,\vec s)\right] \nonumber\\ &+&
{1\over9}\,\sigma_3(\vec\rho,\vec s) \left[\sigma_6(\vec
r_{12},\vec\rho,\vec s) + \sigma_6(\vec r_{13},\vec\rho,\vec
s)\right] \nonumber\\ &+& {1\over9}\,\, \left[\sigma_5(\vec
r_{12},\vec\rho,\vec s)\, \sigma_6(\vec r_{12},\vec\rho,\vec s) +
\sigma_5(\vec r_{13},\vec\rho,\vec s)\, \sigma_6(\vec
r_{13},\vec\rho,\vec s)\right] \nonumber\\ &+& {5\over18}\,
\left[\sigma_5^2(\vec r_{12},\vec\rho,\vec s) + \sigma_5^2(\vec
r_{13},\vec\rho,\vec s) + \sigma_6^2(\vec r_{12},\vec\rho,\vec s)
+ \right.\nonumber\\ &+& \left. \sigma_6^2(\vec
r_{13},\vec\rho,\vec s)\right] + {2\over9}\,\sigma_5(\vec
r_{12},\vec\rho,\vec s)\, \sigma_5(\vec r_{13},\vec\rho,\vec s)
\nonumber\\ &-& {1\over3}\,\sigma_6(\vec r_{12},\vec\rho,\vec s)\,
\sigma_6(\vec r_{13},\vec\rho,\vec s) - {1\over6}\,
\left[\sigma_5(\vec r_{12},\vec\rho,\vec s)\, \sigma_6(\vec
r_{13},\vec\rho,\vec s) \right.\nonumber\\ &+& \left.
\sigma_5(\vec r_{13},\vec\rho,\vec s)\, \sigma_6(\vec
r_{12},\vec\rho,\vec s)\right] \ . \label{435}
 \eeqn

The dipole cross sections $\sigma_k$ ($k=3,4,5,6$) vanish linearly
in $\vec s$ as $\vec s\to 0$. Since the mean value of $s$,
controlled by the distribution function $\Psi_{Pr}(\vec\rho,\vec
s)$, is small, $\la\vec s^{\,2}\ra\sim 1/m_Q^2$, one can make an
expansion,
 \beq
\sigma(\vec r+\vec s_{1,2}) - \sigma(\vec r) \approx \vec
s_{1,2}\cdot\vec\nabla\, \sigma(\vec r)\ , \label{440}
 \eeq
 where $\vec s_1=\beta\vec s$, $\vec s_2=-(1-\beta)\vec s$,
and $\vec r$ is one of the distances $\vec\rho,\ \vec r_{ik}$,
etc.

\section{Saturation of unitarity and breakdown of QCD
factorization}\label{survival}

Factorization assumes that the hard interaction of partons and subsequent
hadronization proceed independently of the soft spectator partons in the
beam and target. This cannot be true for diffraction associated with a large
rapidity gap (LRG). Indeed, the short range hard interaction of partons
guarantees an overlap of the colliding hadrons, large impact parameters do
not contribute. It is known from data that for such near central collisions
unitarity is almost saturated \cite{k3p}, i.e. the chance for colliding
hadrons to escape without soft inelastic interactions which terminate the
LRG is very small.

Here we rely on a simple eikonal model \cite{kps1,bkss,kpst-dy}.
The absorptive corrections to the hard diffractive amplitude lead
to a suppression factor,
 \beq
A_{if}(b)\Rightarrow A_{if}(b)\, \left[1-{\rm Im}
f^{pp}_{el}(b)\right]\ ,
 \label{442}
 \eeq
 where $f^{pp}_{el}(b)$ is the partial elastic amplitude. We assume
a Gaussian shape for the elastic and diffraction amplitudes. After
squaring the amplitude Eq.~(\ref{442}) and integrating over impact
parameter we arrive at the following suppression factor for the
diffractive cross section Eq.~(\ref{410}),
 \beqn
K&=&\left\{1-{1\over{\pi}}\,\frac{\sigma^{pp}_{tot}(s)}
{B_{sd}(s)+2B^{pp}_{el}(s)} \right.\nonumber\\ &+& \left.
 \frac{1}{(4\pi)^2}\,
\frac{\left[\sigma^{pp}_{tot}(s)\right]^2} {B^{pp}_{el}(s)
\left[B_{sd}(s)+B^{pp}_{el}(s) \right]}\right\}\ .
 \label{444}
 \eeqn
 Here the elastic slope depends on energy as
$B^{pp}_{el}(s)=B^0_{el}+2\,\alpha^\prime_{\Pom}\, \ln(s/s_0)$
with $B^{0}_{el}=7.5\GeV^{-2}$, $s_0=1\GeV^2$. The slope of
single-diffractive hard cross section can be estimated as,
$B_{sd}(s)\approx \la r_{ch}^2\ra/3 +2\,\alpha^\prime_{\Pom}$,
where the proton mean charge radius squared is $\la
r_{ch}^2\ra=0.8\fm^2$.

A more accurate estimate needs a detailed information about the
transverse structure of different Fock states. Such information is
very much model dependent. Upon reaching the unitarity limit
(Froissart bound) the fraction of diffractive events is expected
to vanish as $1/\ln(s)$ \cite{brazil}. How soon it may happen
depends on specific models. For instance if gluons form dense
spots inside hadrons, those spots can approach the unitarity limit
much faster than the whole hadron-hadron scattering amplitude.
Such black spots will suppress diffractive gluonic reactions
(heavy flavors, triple-Pomeron term, etc.) much more than is
suggested by Eq.~(\ref{444}). Therefore, the predicted energy
dependence of the survival probability Eq.~(\ref{444}) might be
quite wrong and the diffractive cross section at the LHC energy
may be overestimated.

Gribov corrections \cite{gribov} were introduced into the survival
probability by means of a multi-channel treatment of the unitarity
corrections in \cite{levin1,levin2,ryskin,kaidalov}. The resulting
suppression factor is close to the eikonal one Eq.~(\ref{444}),
differing by less than $10\%$. As an effective description, one
can also incorporate the unitarity effects into the renormalized
Pomeron flux \cite{dino}. This description has the advantage of
simplicity.

\section{Numerical results}\label{results}

Now we are in a position to perform the integrations in
Eqs.~(\ref{390}) and (\ref{410}). Integrating the proton
light-cone wave function squared, $|\Psi_{in}(\vec r_i,x_i)|^2$,
over all variables, except one of the $x_i$, one gets the valence
quark distribution function, ${1\over3}[2u_v(x)+d_v(x)]$. Assuming
a Gaussian dependence on $r_i$ and factorized dependence on both
$r_i$ and $x_i$, one can perform the full integration in
(\ref{390}), (\ref{410}) and single out the contributions of the
different mechanisms. We rely on the phenomenological dipole cross
section which has a saturated shape,
 \beq
\sigma(r,\tilde x)=\sigma_0 \left[1-e^{-r^2/R_0^2(\tilde
x)}\right]\ , \label{450}
 \eeq
 where the parameters have been fitted to HERA data for the
proton structure function at small $\tilde x$ in Ref.~\cite{gbw}.
Here $\tilde x=4m_Q^2/xs$.

First, we found that the bremsstrahlung mechanism, although
leading twist, is very much suppressed. It contributes only a few
percent of the production mechanism at the energy of RHIC, and
order of magnitude less than that at the energy of LHC. Therefore,
we can safely neglect the bremsstrahlung term.

Second, we can disentangle between the contribution coming from
diffractive excitation of individual valence quarks corresponding
to the upper line of graphs in Fig.~\ref{pp-pr}), and interference
terms indicated by the bottom line of graphs in Fig.~\ref{pp-pr}.
The latter is controlled by the parameter $Z=\la
r_{ch}^2\ra_p/R_0^2(\tilde x)$ where the numerator is the proton
mean charge radius squared, $\la r_{ch}^2\ra_p\approx 0.8\fm^2$,
and the denominator is defined in (\ref{450}). The interference
terms vanish like $1/Z$ at large $Z$. This is because the
diffraction amplitude is proportional to the difference between
the cross sections of fluctuations of different size
\cite{brazil}. However, at $r>R_0$ the cross section levels off
and the interference diffractive amplitudes vanish. This general
feature of diffraction is realized in Eq.~(\ref{440}). At high
energies $R_0(\tilde x)=0.4\fm\,(\tilde x/x_0)^{0.144}$ with
$x_0=3\times10^{-4}$ \cite{gbw}. Thus, for charm production
$R_0\approx0.35\fm$ ($Z=7$) at the energy of RHIC, and
$R_0=0.07\fm$ ($Z=160$) at the energy of LHC. Numerical
calculations using Eq.~(\ref{410}) confirm the smallness of
interference terms, which provide only about $1\%$ of the cross
section.

Thus, with good accuracy we can neglect the interference terms in
the production mechanism. This step considerably simplifies the
further calculations. Indeed, since the diffractive $pp$ cross
section comes out as a sum of diffractive excitations of the
proton constituents, we can add sea quarks and gluons as well,
i.e. make a replacement,
 \beq
\bigl|\Psi_{in}(\vec r_i,x_i)\bigr|^2 \Rightarrow
{1\over3}\left[\sum\limits_q q(x)+\bar q(x) +
{81\over16}\,g(x)\right]\ . \label{452}
 \eeq
 We remind the reader that diffractive excitation of a gluon
should be calculated differently from that of a quark, as
described in Sect.~\ref{gluons}.

Next, we should specify the QCD couplings $\alpha_s$ in
(\ref{270}), (\ref{314}). One of them, $\alpha_s(1/s)$ corresponds
to the hard scale of the reaction, $s\sim1/2m_Q$. We use one loop
approximation with three, four and five flavors for the production
of charm, beauty and top respectively. The coupling in (\ref{270})
and (\ref{314}), $\alpha_s(1/rho)$, should be taken on a soft
scale, $\la1/\rho\ra$. As we have just explained, very large
distances $\rho$ are suppressed, since the saturated dipole cross
section levels off and is independent of $\rho$. Therefore the
typical scale for this coupling is controlled by the saturation
scale $Q_s=2/R_0(x)$. Since it partially cover rather low values
of $Q_s$, the problem of infra-red behavior of $\alpha_s$ may
become an issue. We freeze the coupling at the critical value
\cite{gribov-conf} $\alpha_s=3\pi/4(1-\sqrt{2/3})$ (see discussion
in \cite{dokshitzer,k3p}.

\begin{figure}[h]
\begin{center}
 \includegraphics[width=8cm]{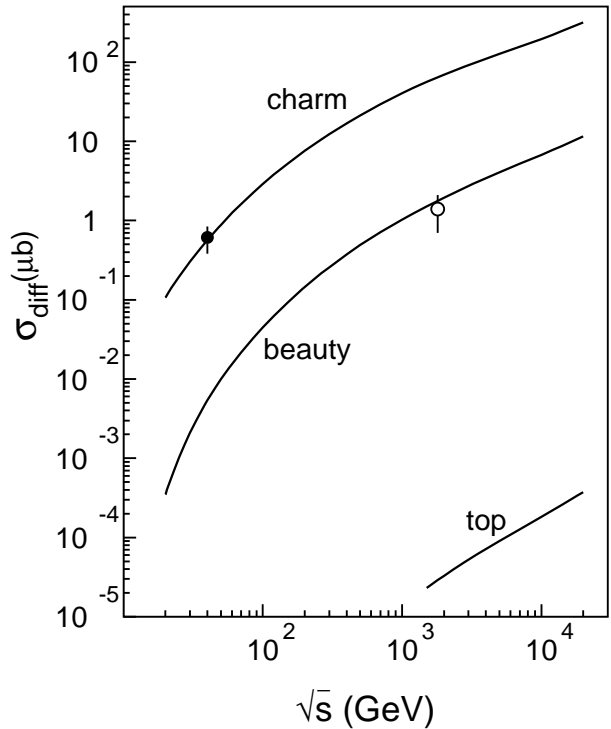}
 \end{center}
 {\caption[Delta]{Cross section of diffractive production of heavy
flavors as function of energy. The experimental points are the
results of the E690 \cite{e690} and CDF \cite{cdf} experiments.}
 \label{s-dep}}
 \end{figure}

The results for the cross section of diffractive production of
charm, beauty and top, $p+p\to \bar QQX+p$, are plotted as
function of energy in Fig.~\ref{s-dep}.
 To be compared with available data (see next section) charm diffractive cross
section is integrated over $x_F>0.85$, and beauty over $x_F>0.9$
(same for top). All the cross sections steadily rise with energy.
The cross sections of charm and beauty production differ by about
an order of magnitude what confirms the expected leading twist
behavior $1/m_Q^2$.

\begin{figure}[h]
\begin{center}
 \includegraphics[width=7.5cm]{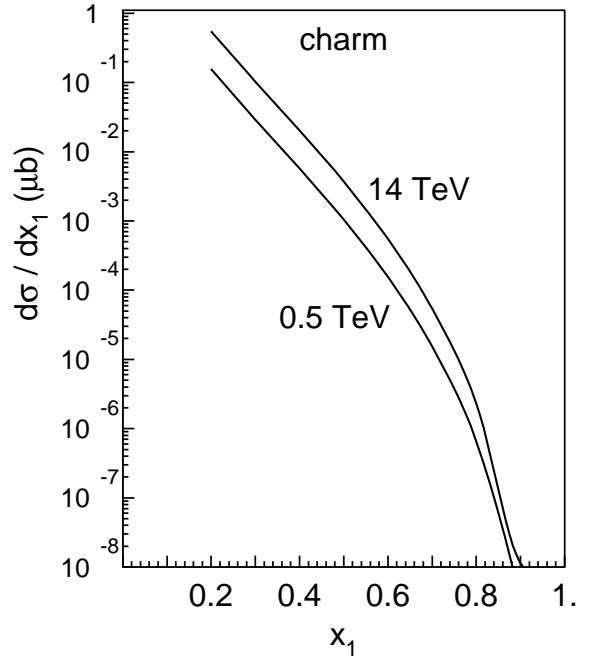}
 \end{center}
 {\caption[Delta]{The cross section of diffractive excitation of a proton
with charm production as function of fraction $x_1$ of the initial
proton momentum carried by the charm quark. The cross section is
shown at the energies of RHIC and LHC.}
 \label{c-x1}}
 \end{figure}

We also calculated the $x_1$ distribution of a diffractively
produced charm quark by integrating over all other variables.
$x_1=p_c^+/p_p^+$ is the ratio of plus components of the produced
$c$-quark and the incoming proton. The results are shown in
Fig.~\ref{c-x1} at RHIC and LHC energies.

 Notice that to be compared with data (unavailable so far) for production
of charmed mesons, this result has to be corrected for the
fragmentation $c\to D$ which is poorly known. The resulting
behavior at $x_1\to 1$ should obey the end-point behavior dictated
by Regge. Therefore we expect it to be less steep than what is
plotted in Fig.~\ref{c-x1}. One may wonder: a convolution with the
fragmentation function $c\to D$ may only result in a steeper fall
off at $x_1\to 1$, how can it become less steep? The answer is:
the convolution procedure is incorrect, QCD factorization badly
fails at $x_1\to 1$. The usual fragmentation function measured,
say, in $e^+e^-$ annihilation, corresponds to a fast $c$-quark
producing a jet and picking up a slow light quark from vacuum to
form a $D$-meson. In hadronic collisions at large $x_1$
hadronization occurs differently: a fast projectile light quark
picks up a slow $c$-quarks produced perturbatively.
Correspondingly, in the case of diffractive production of a heavy
flavored baryon a leading projectile diquark can pick up the heavy
quark.

Notice also that $x_1$ has a bottom bound imposed by the
kinematics of diffraction, $x_1>4m_Q^2/(1-x_F)s$, where $x_F$ is
the Feynman variable of the recoil proton in $pp\to Xp$. In order
to comply with available data (see next section) we integrate over
$x_F>0.85$ for charm (also top), and $x_F>0.9$ for beauty.

\begin{figure}[h]
\begin{center}
 \includegraphics[width=7.5cm]{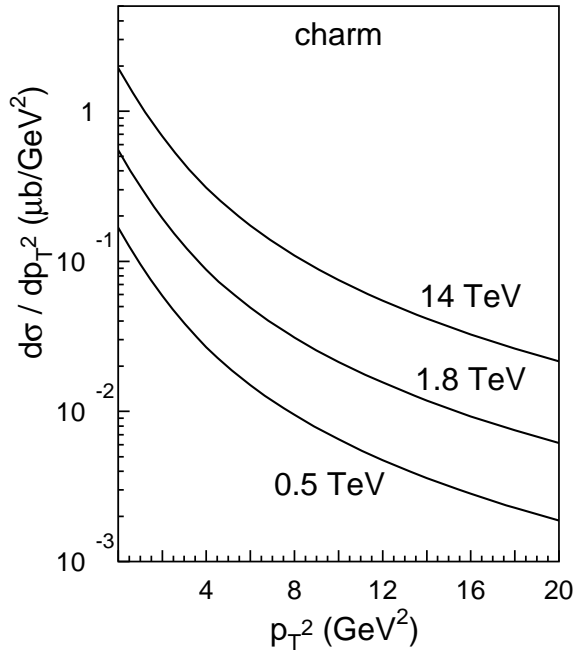}
 \end{center}
 {\caption[Delta]{The $p_T$ dependence of the cross section of
diffractive excitation of charm.}
 \label{c-pt}}
 \end{figure}

\begin{figure}[h]
\begin{center}
 \includegraphics[width=7cm]{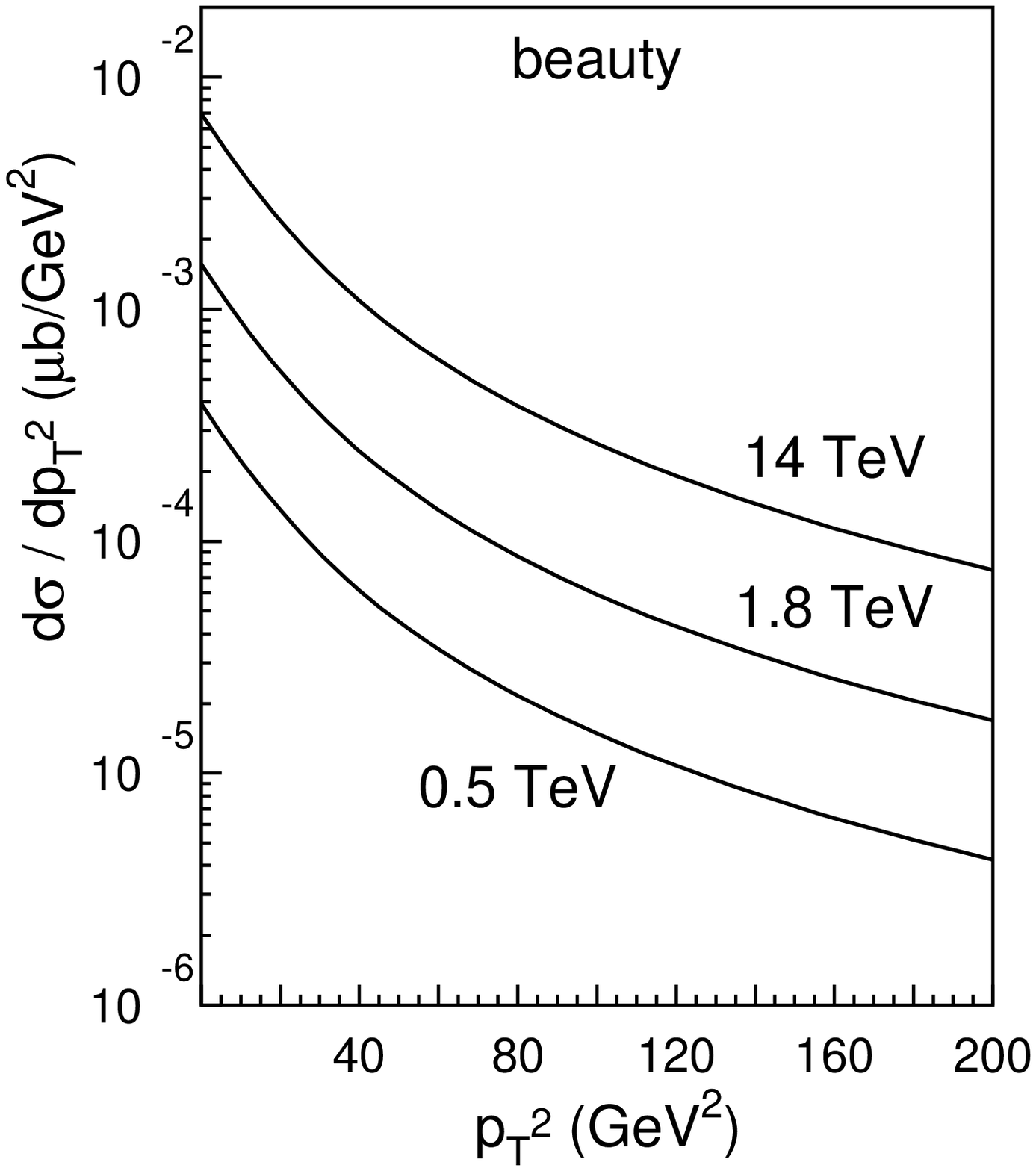}
 \end{center}
 {\caption[Delta]{The $p_T$ dependence of the cross section of
diffractive excitation of beauty.}
 \label{b-pt}}
 \end{figure}

\begin{figure}[h]
\begin{center}
 \includegraphics[width=7cm]{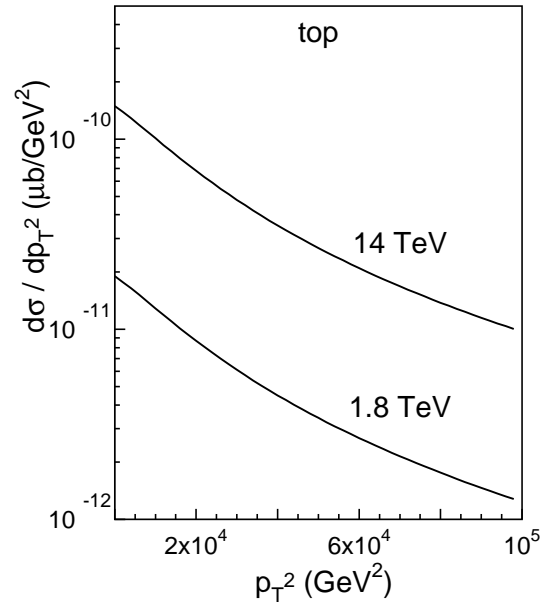}
 \end{center}
 {\caption[Delta]{The $p_T$ dependence of the cross section of
diffractive excitation of top.}
 \label{t-pt}}
 \end{figure}

Our results for transverse momentum distribution of diffractively
produced quarks are presented in Figs.~(\ref{c-pt})-(\ref{t-pt})
for different heavy flavors and energies.

There $p_T$ distributions hardly correlate with $x_1$ of the heavy
quark, what is quite different from the usual sea-gull effect. We
remind, however, that this is not the usual factorization based
hadronization. In this case a fast projectile quark-spectator
picks up a slow heavy flavor. Therefore, the transverse momentum
of the produced heavy flavored meson is mainly controlled by the
transverse momentum of the light spectator.

To conclude this section, we should comment on the accuracy of
performed calculations. The main uncertainty seems to be related
to the absorptive (unitarity) corrections. Comparing different
models, the difference is not dramatic, of the order of $10\%$,
with a probability factor $K=0.14$ at the Tevatron energy.
However, all those models may miss the specific dynamics of
interaction discussed in Sect.~\ref{survival} and overestimate
diffraction at the LHC energy by much more than $10\%$. The next
theoretical uncertainty is related to the choice of heavy quark
masses. We used $m_c=1.5\GeV$, $m_b=4.7\GeV$ and $m_t=175\GeV$.
Our diffractive cross sections approximately scale as $1/m_Q^2$.
The cross section also depends on the behavior of the QCD coupling
in the infrared limit. We freeze the coupling at the critical
value $\alpha_s\leq 0.4$. A larger value would lead to a
corresponding increase of the cross section.

\section{Data}\label{data}

\subsection{Diffractive production of charm}\label{charm}

Data for diffractive production of heavy flavors are scarce. Three
experiments searched for open charm production in diffractive
events. Diffractive production of $\Lambda_c$ was first detected
in experiment \cite{giboni} at $\sqrt{s}=63\GeV$ at ISR. The
reported cross section $\sigma(pp\to\Lambda_cXp) = 10-40\mub$ is
quite above our calculations.  However, one should take this
experimental result with precaution, since it also considerably
exceeds the results of a later measurement (see below). The same
experiment overestimated the total cross section of inclusive
charm production by more than an order of magnitude (see
discussion in \cite{e690}).

A stringent upper limit for diffractive charm production at
$\sqrt{s}=40\GeV$ was imposed by the E653 experiment at Fermilab
\cite{e653}, which found the cross section to be smaller than
$26\mub$ for proton-silicon collisions. These measurements
included events with the nucleus remaining intact, as well as
fragmenting. The nuclear effects are controlled mostly by the
survival probability of the large rapidity gap during propagation
of projectile soft spectator partons through the nucleus.
Therefore, only nuclear periphery contributes and nuclear effects
are rather small. According to calculations in \cite{kps1,kps2},
the cross sections on silicon and free proton are similar, so one
can apply the measured upper bound for $pp$ collisions as well.
Our calculations are well below this bound.

An accurate measurement of the cross section of diffractive
production of $D^*$-mesons was performed in the E690 experiment at
Fermilab at $\sqrt{s}=40\GeV$ and $x_F>0.85$ \cite{e690}. The
result for the cross section of $\bar cc$ production based on the
most accurate data for production of $D^{*-}$ is $\sigma(p+p\to
\bar ccX+p) = [0.61 \pm 0.12(stat) \pm 0.11(syst)] \mub$. This
cross sections is integrated over $x_F>0.85$. Plotted in
Fig.~\ref{s-dep} it agrees well with our calculations.

\subsection{Diffractive production of beauty}\label{beauty}

An upper limit for the cross section of diffractive beauty
production at $\sqrt{s}=630\GeV$ was established by the UA1
experiment, with a large theoretical uncertainty: $\sigma(\bar
pp\to bX+p/\bar p)\leq[0.6-1.2]\mub$. The corresponding upper
limit for the fraction of diffractive production relative to
inclusive production of beauty was found to be $R_{\bar
bb}<[3.1-6.2]\%$.

The first observation of diffractive beauty production was done by
the CDF collaboration \cite{cdf} at $\sqrt{s}=1800\GeV$ and
$x_F>0.9$.  Unfortunately the cross section suffers large
theoretical uncertainties. The Monte Carlo codes used for the
evaluation of the gap acceptance were based on factorization,
which is broken for diffraction. Depending on the assumed
distributions of quarks and gluons in the Pomeron, the fraction of
diffractive events with $x_F>0.9$ ranges from $R_{\bar
bb}=[0.62\pm0.19(stat)\pm0.16(syst)]\%$ to
$[1.18\pm0.36(stat)\pm0.27(syst)]\%$ \cite{phd}.  From these
results one can probably conclude only that the ratio is of the
order of $1\%$ with the error of the order of $0.5\%$. We use this
estimate for further comparison with our calculations.

It also worth mentioning that the electrons from $b$-decay are
detected with large transverse momenta $p_T>9.5\GeV$ \cite{cdf}.
As far as diffraction this is leading twist, and it should have a
$p_T$-dependence similar to that in inclusive production. Then the
experimental $p_T$-cut should not affect much the
diffraction-to-inclusive ratio.

To get the diffractive cross section one needs to know the
inclusive cross section of beauty production which unfortunately
has not been measured at this energy. One can rely on the result
of the UA1 experiment at $\sqrt{s}=630\GeV$, which is $\sigma(\bar
pp\to b+X)=[19.3\pm7(exp)\pm9(theor)]\mub$, and also on data at
lower energies \cite{e789,e771,hera-b} and a theoretical
extrapolation. Two theoretical approaches explain reasonably well
data for inclusive production of beauty, the NLO parton model
\cite{nlo} and the light-cone dipole description \cite{kr}.
Although the absolute predictions differ by about a factor of 2,
they predict the same energy dependence. With this energy
dependence we extrapolated the result of the UA1 at
$\sqrt{s}=630\GeV$ to the Tevatron energy $\sqrt{s}=1800\GeV$, and
found $\sigma(\bar pp\to\bar bbX)=[103.5\pm86]\mub$ (we added
linearly the experimental and theoretical uncertainties).

This result is quite uncertain because of the large theoretical
error and of the long energy interval for extrapolation. Another
possibility to make an estimate is to use CDF data
\cite{cdf-inclusive} for inclusive $b$-quark production at
$\sqrt{s}=1960\GeV$ in the rapidity interval $|y|<0.6$,
$\sigma=[17.6\pm0.4(stat)^{+2.5}_{-2.3}(syst)]\mub$, and to
extrapolate it to other rapidities. A rough estimate would be to
assume the same production rate for the whole rapidity interval
$\Delta y=2\ln(\sqrt{s}/m_{bT})$. We take the mean transverse mass
$m_{bT}^2=m_b^2+\la p_T^2\ra\approx 2m_b^2$. The last small
correction is to scale these numbers according to the theoretical
energy dependence \cite{kr},
$\sigma(\sqrt{s}=1960GeV)/\sigma(\sqrt{s}=1800GeV)=0.95$.
Eventually, we arrive at the estimate, $\sigma(\bar pp\to
b+X)=[139.7\pm3(stat)^{+20}_{-18}(syst)]\mub$.

Alternatively, one can rely on theoretical predictions for the
inclusive cross section done within the dipole approach. Usually
such calculations are pretty accurate, since are based on the
phenomenological dipole cross section fitted to DIS data, and
include all higher order corrections and higher twists. In
particular, it describes quite well the available data for
inclusive production of charm. The predicted inclusive cross
section of inclusive beauty production at $\sqrt{s}=1800\GeV$ is
$140\mub$, which agrees well with the above extrapolated
experimental cross sections.

Relying on this theoretically predicted inclusive cross section
and experimentally measured \cite{cdf} fraction $R_{\bar bb}$ of
diffractively produced beauty, we finally arrive at the
diffractive cross section, $\sigma_{sd}(\bar pp\to
bXp)=[1.4\pm0.7]\mub$. This value is plotted in Fig.~\ref{s-dep}
in comparison with our calculations. Still, one should remember
that both the value and error are subject to considerable
uncertainties, in particular theoretical ones.

\section{Summary}\label{summary}

The main results of this paper can be summarized as follows.

\begin{itemize}

\item Novel leading twist mechanisms of diffractive excitation of heavy flavors in
hadronic collision are proposed and calculated. Factorization
leading to higher twist diffraction is badly broken.

\item Two mechanisms of heavy flavor production are identified (see
Fig.~\ref{graph}). One, called bremsstrahlung, is similar to the
Drell-Yan mechanism of radiation of a heavy dilepton, but includes
also interaction of the radiated virtual gluon with the target.
Another mechanism, called production, involves also interaction of
the heavy quarks with the target. Diffraction excitation of $\bar
QQ$ in a separate parton by the bremsstrahlung mechanism is higher
twist and can be neglected. The leading twist excitation of
projectile quarks or gluons is possible only via the productive
mechanism.

\item The presence of spectator partons in the projectile hadron opens new
possibilities of interactions, and the bremsstrahlung mechanism
becomes leading twist as well. Quantitatively, however, it is
still a small part of the cross section. The dominant contribution
comes from diffractive excitation of a separate projectile quark
or gluon via the production mechanism.

\item Available data for diffractive production of charm and beauty agree with our
calculations. The leading twist dependence on the quark mass is
also confirmed.

\item Our results for heavy flavors allow straightforward application to
high-$p_T$ jets. The same leading twist production mechanism
explains the observed independence of hard scale of the
diffraction-to-inclusive ratio for di-jet production
\cite{cdf-pt}. Numerical calculations and comparison with data
will be done elsewhere.

\end{itemize}

\begin{acknowledgments}

We are grateful to Stan Brodsky for inspiring discussions which motivated us to perform
the present calculations. We are also thankful to Hans-J\"urgen Pirner for very useful
discussions and to Dino Goulianos for providing information about CDF data.  We thank
Genya Levin for his comments clarifying the results of \cite{levin-coh}.  This work was
supported in part by Fondecyt (Chile) grants, numbers 1050519, 1050589 and 7050175, and
by DFG (Germany)  grant PI182/3-1.

\end{acknowledgments}


 \def\appendix{\par
 \setcounter{section}{0}
\setcounter{subsection}{0}
 \def\thesection{Appendix \Alph{section}}
\def\thesubsection{\Alph{section}.\arabic{subsection}}
\def\theequation{\Alph{section}.\arabic{equation}}
\setcounter{equation}{0}}

 \appendix

\section{Classification of the amplitudes}
\label{classified} \setcounter{equation}{0}

\subsection{Transverse polarization}
\label{1.T}

The five amplitudes corresponding to the Feynman graphs in
Fig.~\ref{graphs}, with transversely polarized gluons radiated by
the projectile quark have the following form,\\
 \beqn
M_1^T \hspace*{-1mm} &=& \hspace*{-1mm}
 \frac{\sum\limits_\lambda\Gamma_1^\lambda(-\vec
k)\,\Gamma_2^\lambda(\vec\kappa)\,
\tau_a^{q_1}\,\tau_a^{q_1}\,\tau^Q_b\ f_a(\vec q)}
{[M^2(1-\alpha)+\alpha^2m_q^2+k^2][m_Q^2+\kappa^2]};
\label{1.10}\\ M_2^T \hspace*{-1mm} &=& \hspace*{-1mm}
\frac{\sum\limits_\lambda\Gamma_1^\lambda(\vec\pi)\,
\Gamma_2^\lambda(\vec\kappa)\,
\tau_b^{q_1}\,\tau_a^{q_1}\,\tau^Q_b\ f_a(\vec q)}
{[M^2(1-\alpha)+\alpha^2m_q^2+\pi^2][m_Q^2+\kappa^2]};
\label{1.20}\\ M_3^T \hspace*{-1mm} &=& \hspace*{-1mm}
\frac{\sum\limits_\lambda\Gamma_1^\lambda(\vec p_2)\,
\Gamma_2^\lambda(\vec\kappa)\, if_{abc}\, \tau_c^{q_1}\,\,\tau_b\
f_a(\vec q)} {[p_2^2+\alpha^2m_q^2][m_Q^2+\kappa^2]};
\label{1.30}\\ M_4^T \hspace*{-1mm} &=& \hspace*{-1mm}
\frac{\sum\limits_\lambda\Gamma_1^\lambda(\vec p_2)\,
\Gamma_2^\lambda(\vec\kappa-\beta\vec q)\,
\tau_b^{q_1}\,\tau^Q_b\,\tau^Q_a\ f_a(\vec q)}
{[p_2^2+\alpha^2m_q^2][m_Q^2+(\vec\kappa-\beta\vec q)^2
+\beta(1-\beta)Q^2]};\nonumber\\ \label{1.40}\\ M_5^T
\hspace*{-1mm} &=& - \frac{1}{p_2^2+\alpha^2m_q^2} \hspace*{-1mm}
\nonumber\\&\times& \hspace*{-1mm}
\frac{\sum\limits_\lambda\Gamma_1^\lambda(-\vec p_2)\,
\Gamma_2^\lambda[\vec\kappa+(1-\beta)\vec q]\,
\tau_b^{q_1}\,\tau^Q_a\,\tau^Q_b\ f_a(\vec q)}
{m_Q^2+[\vec\kappa+(1-\beta)\vec q]^2
+\beta(1-\beta)Q^2}.\nonumber\\ \label{1.50}
 \eeqn
 Here, $f_a(\vec q)$ is the amplitude for emission of a gluon with color
index $a$ and transverse momentum $\vec q$ by the target proton;
 \beqn
\Gamma_1^{\lambda}(\vec\pi) &=&
\chi_f^\dagger\,\Bigl\{(2-\alpha)(\vec e_\lambda\cdot\vec\pi) +
i\alpha[\vec\sigma\times\vec e_\lambda]\cdot\vec\pi \nonumber\\
&+& i\alpha^2m_q\, [\vec\sigma\times\vec e_\lambda]\cdot\vec n
\Bigr\}\,\chi_{in}; \nonumber\\ \Gamma_2^{\lambda}(\vec\kappa) &=&
\phi^\dagger\,\Bigl\{m_1(\vec\sigma\cdot\vec e_\lambda)+
(1-2\beta)(\vec\sigma\cdot\vec n) (\vec e_\lambda\cdot\vec\kappa)
\nonumber\\ &+& i[\vec e_\lambda\times\vec n]
\cdot\vec\kappa\Bigr\}\,\bar\phi; \label{1.70}
 \eeqn

Notice that within each of the two groups of amplitudes, $M_1^T$,
$M_2^T$, and $M_4^T$, $M_5^T$, the denominators have similar
structure and these amplitudes may be combined producing
light-cone distribution amplitudes in impact parameter
representation. Only the amplitude $M_3^T$ does not belong to any
of these groups. Nevertheless, it can be splited into two parts.
The first one comes out after multiplying $M_3^T$ by factor,
 \beq
\frac{Q^2}{M^2+Q^2} = \frac{p_2^2+\alpha^2 m_q^2}
{M^2(1-\alpha)+\alpha^2m_q^2+p_2^2}\ . \label{1.80}
 \eeq
 Then the amplitude acquires the denominator of the first group, which we
call bremsstrahlung.  The rest of $M_3^T$, which gets the factor,
 \beq
\frac{M^2}{M^2+Q^2} = \frac{m_Q^2+\kappa^2}
{m_Q^2+\kappa^2+\beta(1-\beta)Q^2}\ , \label{1.90}
 \eeq
 has the structure corresponding to the second group, which
we call production mechanism. This explains the way in which we
classify the amplitudes in Eqs.~(\ref{10})-(\ref{20}).

Using the relation,
 \beq
if_{abc}\tau^{q_1}_c\tau^Q_b=\left[\tau^{q_1}_a\tau^{q_1}_b-
\tau^{q_1}_b\tau^{q_1}_a\right]\tau^Q_b =
(\tau^Q_c\tau^Q_a-\tau^Q_a\tau^Q_c)\tau^{q_1}_c\ , \label{1.100}
 \eeq
 we obtain for the bremsstrahlung and production amplitudes
Eqs.~(\ref{10})-(\ref{20}), respectively,
 \beqn
M^T_{Br}&=&\Bigl\{\left[ \Phi^T_{Br}\bigl(\vec\pi+(1-\alpha)\vec
q,\vec\kappa\bigr)- \Phi^T_{Br}(\vec\pi-\alpha\vec
q,\vec\kappa)\right] \nonumber\\ &\times&
\tau^{q_1}_a\tau^{q_1}_b\tau^Q_b +
\Bigl[\Phi^T_{Br}(\vec\pi,\vec\kappa) \nonumber\\ &-&
\Phi^T_{Br}\bigl(\vec\pi+ (1-\alpha)\vec q,\vec\kappa\bigr)\Bigr]
\tau^{q_1}_b\tau^{q_1}_a\tau^Q_b\Bigr\}\, f_a(\vec q)\
,\nonumber\\ \label{1.110}
 \eeqn
 where
 \beq
\Phi^T_{Br}\bigl(\vec\pi,\vec\kappa\bigr)=
\frac{\sum\limits_\lambda\Gamma_1^\lambda(\vec\pi)\,
\Gamma_2^\lambda(\vec\kappa)}
{\left[M^2(1-\alpha)+\alpha^2m_q^2+\vec\pi^{\,2}\right]
(m_Q^2+\kappa^2)}\ . \label{1.120}
 \eeq
 Correspondingly, for the production mechanism,
 \beqn
M^T_{Pr}&=&\Bigl\{\left[ \Phi^T_{Pr}\bigl(\vec
p_2,\vec\kappa-\beta\vec q\bigr)- \Phi^T_{Pr}(\vec
p_2,\vec\kappa)\right] \tau^{q_1}_b\tau^Q_b\tau^Q_a \nonumber\\
&+& \Bigl[\Phi^T_{Pr}(\vec p_2,\vec\kappa)- \Phi^T_{Pr}\bigl(\vec
p_2,\vec\kappa+(1-\beta)\vec q\bigr)\Bigr] \nonumber\\ &\times&
\tau^{q_1}_b\tau^Q_a\tau^Q_b\Bigr\}\, f_a(\vec q)\ , \label{1.130}
 \eeqn
 where
 \beq
\Phi^T_{Pr}\bigl(\vec p_2,\vec\kappa\bigr)=
\frac{\sum\limits_\lambda\Gamma_1^\lambda(\vec\pi)\,
\Gamma_2^\lambda(\vec\kappa)} {\left(p_2^2+\alpha^2m_q^2\right)
\left[m_Q^2+\kappa^2+\beta(1-\beta)Q^2\right]}\ . \label{1.140}
 \eeq

Notice that the amplitudes Eqs.~(\ref{1.110}),~(\ref{1.130})
vanish in the forward direction, at $\vec q\to0$

Now we can convert the distribution amplitude Eq.~(\ref{1.140} to
impact parameter representation,
 \beq
\Phi^T_{Br(Pr)}(\vec\rho,\vec s) = \frac{1}{(2\pi)^4}\int
d^2\pi\,d^2\kappa\, e^{-i\vec\pi\cdot\vec\rho-
i\vec\kappa\cdot\vec s}\, \Phi^T_{Br(Pr)}(\vec\pi,\vec\kappa)\ ;
\label{1.150}
 \eeq
 \beq
\gamma_a(\vec b) ={1\over2\pi} \int d^2q\, e^{-i\vec q\cdot\vec
b}\, f_a(\vec q)\ . \label{1.160}
 \eeq

Then the amplitudes Eqs.~(\ref{1.110}),~(\ref{1.130}) take the
form,
 \beqn
&&M^T_{Br}= \int d^2b\,d^2\rho\,d^2s\, e^{i\vec q\cdot\vec
b+i\vec\pi\vec\rho+ i\vec\kappa\cdot\vec s}\,
\Phi^T_{Br}(\vec\rho,\vec s) \nonumber\\ &\times&
\Bigl\{\left[\gamma_a\left(\vec b-(1-\alpha)\vec\rho\right) -
\gamma_a(\vec b+\alpha\vec\rho) \right]\,
\tau^{q_1}_a\tau^{q_1}_b\tau^Q_b \nonumber\\ &+&
\left[\gamma_a(\vec b) - \gamma_a(\vec b+\alpha\vec\rho) \right]\,
\tau^{q_1}_b\tau^{q_1}_a\tau^Q_b \Bigr\}\,; \label{1.170}
 \eeqn

 \beqn
&&M^T_{Pr}= \int d^2b\,d^2\rho\,d^2s\, e^{i\vec q\cdot\vec
b+i\vec\pi\vec\rho+ i\vec\kappa\cdot\vec s}\,
\Phi^T_{Pr}(\vec\rho,\vec s) \nonumber\\ &\times&
\Bigl\{\left[\gamma_a\left(\vec b-(1-\alpha)\vec\rho +\beta
s\right) - \gamma_a\left(\vec b-(1-\alpha)\vec\rho\right) \right]
\nonumber\\ &\times& \tau^{q_1}_b\tau^Q_b\tau^Q_a +
\left[\gamma_a\left(\vec b-(1-\alpha)\vec\rho\right)
\right.\nonumber\\&-&\left. \gamma_a\left(\vec
b-(1-\alpha)\vec\rho -(1-\beta)\vec s\right) \right]\,
\tau^{q_1}_b\tau^Q_a\tau^Q_b \Bigr\}\,. \label{1.180}
 \eeqn

\subsection{Longitudinal polarization}
\label{1.L}

The five amplitudes corresponding to the graphs in
Fig.~\ref{graphs}
 \beqn
M_1^L &=& 4\,\tau^{q_1}_a\tau^{q_1}_b\tau^Q_b\,f_a(\vec q)\,
(\chi_f^\dagger\,\chi_{in})\, (\phi^\dagger\,\vec\sigma\cdot\vec
n\,\bar\phi) \nonumber\\ &\times&
\left[\frac{1-\alpha}{M^2(1-\alpha)+\alpha^2 m_q^2 + \kappa^2} +
\frac{1}{M^2}\right]; \nonumber\\ M_2^L &=&
-4\,\tau^{q_1}_b\tau^{q_1}_a\tau^Q_b\,f_a(\vec q)\,
(\chi_f^\dagger\,\chi_{in})\, (\phi^\dagger\,\vec\sigma\cdot\vec
n\,\bar\phi) \nonumber\\ &\times&
\left[\frac{1-\alpha}{M^2(1-\alpha)+\alpha^2 m_q^2 + \kappa^2} +
\frac{1}{M^2}\right]; \nonumber\\ M_3^L &=&
4if_{abc}\,\tau^{q_1}_c\tau^Q_b\,f_a(\vec q)\,
(\chi_f^\dagger\,\chi_{in})\, (\phi^\dagger\,\vec\sigma\cdot\vec
n\,\bar\phi) \nonumber\\ &\times& \left[{1\over Q^2}-{1\over
M^2}\right]; \nonumber\\ M_4^L &=&
4\,\tau^{q_1}_b\tau^Q_b\tau^Q_a\,f_a(\vec q)\,
(\chi_f^\dagger\,\chi_{in})\, (\phi^\dagger\,\vec\sigma\cdot\vec
n\,\bar\phi) \nonumber\\ &\times&
\left[\frac{\beta(1-\beta)}{m_Q^2+ (\vec\kappa-\beta\vec q)^2 +
Q^2\beta(1-\beta)} - \frac{1}{Q^2}\right]; \nonumber\\ M_5^L &=&
4\,\tau^{q_1}_b\tau^Q_a\tau^Q_b\,f_a(\vec q)\,
(\chi_f^\dagger\,\chi_{in})\, (\phi^\dagger\,\vec\sigma\cdot\vec
n\,\bar\phi) \nonumber\\ &\times&
\left[\frac{\beta(1-\beta)}{m_Q^2+ \bigl(\vec\kappa-(1-\beta)\vec
q\bigr)^2 + Q^2\beta(1-\beta)} - \frac{1}{Q^2}\right]. \nonumber\\
\label{1.190}
 \eeqn

Summing up these amplitudes, the terms proportional to $1/Q^2$ and
$1/M^2$ cancel. The rest, the amplitudes $M_i^L$ ($i=1,2,4,5$)
without these terms, which we denote $\widetilde{M}^L_i$, can be
grouped in a way to create light-cone distribution amplitudes. To
reach this goal we introduce an additional amplitude which is
identical to zero,
 \beq
\widetilde M^L_3 = \widetilde M^L_{3,1} - \widetilde M^L_{3,2}
\equiv 0\ , \label{1.200}
 \eeq
 where
 \beqn
\widetilde M_{3,1}^L &=& 4if_{abc}\,\tau^{q_1}_c\tau^Q_b\,f_a(\vec
q)\, (\chi_f^\dagger\,\chi_{in})\,
(\phi^\dagger\,\vec\sigma\cdot\vec n\,\bar\phi) \nonumber\\
&\times& \frac{1-\alpha}{M^2(1-\alpha)+m_q^2+p_2^2}\ ; \nonumber\\
\widetilde M_{3,2}^L &=& 4if_{abc}\,\tau^{q_1}_c\tau_b\,f_a(\vec
q)\, (\chi_f^\dagger\,\chi_{in})\,
(\phi^\dagger\,\vec\sigma\cdot\vec n\,\bar\phi) \nonumber\\
&\times& \frac{\beta(1-\beta)}{m_Q^2+\kappa^2 + Q^2\beta(1-\beta)}
\label{1.210}
 \eeqn

Now we are in the position to group the longitudinal amplitudes in
a similar form to the transverse ones, getting the light-cone
distribution amplitudes corresponding to the bremsstrahlung and
production mechanisms,
 \beqn
M^L_{Br}&=&\widetilde M_1 + \widetilde M_2 + \widetilde M_{3,1}^L\
; \nonumber\\ M^L_{Pr}&=&\widetilde M_4 + \widetilde M_5 -
\widetilde M_{3,2}^L\ . \label{1.220}
 \eeqn
These amplitudes have the form,
 \beqn
&&M^L_{Br}= \nonumber\\ &&\Bigl\{\left[\Phi^L_{Br}\bigl(
\vec\pi+(1-\alpha)\vec q,\vec\kappa\bigr)- \Phi^L_{Br}\bigl(
\vec\pi-\alpha\vec q,\vec\kappa\bigr)\right] \tau^{q_1}_a
\tau^{q_1}_b \tau^Q_b \nonumber\\ &+&
\Bigl\{\left[\Phi^L_{Br}\bigl( \vec\pi,\vec\kappa\bigr)-
\Phi^L_{Br}\bigl( \vec\pi+(1-\alpha)\vec q,\vec\kappa\bigr)\right]
\tau^{q_1}_b \tau^{q_1}_a \tau^Q_b\Bigr\} f_a(\vec q); \nonumber\\
\label{1.230}
 \eeqn
 \beqn
&&M^L_{Pr}= \nonumber\\ &&\left[\Phi^L_{Pr}\bigl( \vec
p_2,\vec\kappa-\beta\vec q\bigr)- \Phi^L_{Br}\bigl(\vec
p_2,\vec\kappa\bigr)\right] \tau^{q_1}_b \tau^Q_b \tau^Q_a
\nonumber\\ &+& \left[\Phi^L_{Br}\bigl(\vec p_2,\vec\kappa\bigr)-
\Phi^L_{Br}\bigl(\vec p_2,\vec\kappa+(1-\beta)\vec q\bigr)
\right]\tau^{q_1}_b \tau^{q_1}_a \tau^Q_b\Bigr\} f_a(\vec q),
\nonumber\\ \label{1.240}
 \eeqn
 where the longitudinal distribution amplitudes $\Phi^L$ read,
 \beqn
\Phi^L_{Br}(\vec\pi,\vec\kappa) &=& \frac{4(1-\alpha)
(\chi_f^\dagger\,\chi_{in})\, (\phi^\dagger\,\vec\sigma\cdot\vec
n\,\bar\phi)} {M^2(1-\alpha)+\alpha^2m_q^2+\pi^2}; \nonumber\\
\Phi^L_{Pr}(\vec\pi,\vec\kappa) &=& \frac{4\beta(1-\beta)
(\chi_f^\dagger\,\chi_{in})\, (\phi^\dagger\,\vec\sigma\cdot\vec
n\,\bar\phi)} {m_Q^2+\kappa^2+Q^2\beta(1-\beta)}. \nonumber\\
\label{1.250}
 \eeqn

\section{Useful algebras}
\label{algebra} \setcounter{equation}{0}

Average of a product of arbitrary functions $A(\tau),\ B(\tau),\
C(\tau)$ of Gell-Mann matrices over the proton wave function has
the general form,
 \beqn
&&\left\la\, A\bigl(\tau^{q_1}\bigr)\, B\bigl(\tau^{q_1}\bigr)\,
C\bigl(\tau^{q_1}\bigr)\, \right\ra_N \nonumber\\ &=&
{1\over6}\,\Bigl\{\tr A(\tau)\,\tr B(\tau)\, \tr C(\tau)
\nonumber\\ &+& \tr\bigl[A(\tau)\,B(\tau)\,C(\tau)\bigr] +
\tr\bigl[A(\tau)\,C(\tau)\,B(\tau)\bigr] \nonumber\\ &-& \tr
A(\tau)\,\tr\bigl[B(\tau)\,C(\tau)\bigr]- \tr
B(\tau)\,\tr\bigl[A(\tau)\,C(\tau)\bigr] \nonumber\\ &-& \tr
C(\tau)\,\tr\bigl[A(\tau)\,B(\tau)\bigr] \Bigr\} \label{2.10}
 \eeqn

More relations for $\tau$-matrices,
 \beqn
\tr\tau^Q_a &=& 0;\ \ \ \ \
\tr\bigl[\tau_a\tau_b\bigr]={1\over2}\,\delta_{ab}; \nonumber\\
\tr\bigl[\tau_a\tau_b\tau_c\bigr] &=& {1\over4}\, h_{abc},\ \ \
h_{abc}= d_{abc} + if_{abc}; \nonumber\\
\tr\bigl[\tau_a\tau_b\tau_c\tau_d\bigr] &=&
{1\over12}\,\delta_{ab}\,\delta_{cd}+ {1\over8}\,h_{abe}\,h_{cde};
\label{2.20}
 \eeqn

 \beqn
\tau_a\,\tau_b\,\tau_a &=& -{1\over6}\,\tau_b; \nonumber\\
\tau_a\,\tau_b\,\tau_c\,\tau_a &=& {1\over4}\,\delta_{bc} -
{1\over6}\,\tau_b\,\tau_c\ ; \label{2.30}
 \eeqn

 \beqn
f_{akl}f_{bkl} &=& 3\,\delta_{ab}; \nonumber\\ d_{akl}d_{bkl} &=&
{5\over3}\,\delta_{ab}; \nonumber\\ f_{iak}f_{kbl}f_{lci} &=&
-{3\over2}\,f_{abc}; \nonumber\\ d_{iak}f_{kbl}f_{lci} &=&
-{3\over2}\,d_{abc}; \nonumber\\ d_{iak}d_{kbl}f_{lci} &=&
{5\over6}\,f_{abc}; \nonumber\\ d_{iak}d_{kbl}d_{lci} &=&
{1\over2}\,d_{abc}. \label{2.40}
 \eeqn

\end{document}